\newdimen\tableauside\tableauside=1.0ex
\newdimen\tableaurule\tableaurule=0.4pt
\newdimen\tableaustep
\def\phantomhrule#1{\hbox{\vbox to0pt{\hrule height\tableaurule
width#1\vss}}}
\def\phantomvrule#1{\vbox{\hbox to0pt{\vrule width\tableaurule
height#1\hss}}}
\def\sqr{\vbox{%
  \phantomhrule\tableaustep

\hbox{\phantomvrule\tableaustep\kern\tableaustep\phantomvrule\tableaustep}%
  \hbox{\vbox{\phantomhrule\tableauside}\kern-\tableaurule}}}
\def\squares#1{\hbox{\count0=#1\noindent\loop\sqr
  \advance\count0 by-1 \ifnum\count0>0\repeat}}
\def\tableau#1{\vcenter{\offinterlineskip
  \tableaustep=\tableauside\advance\tableaustep by-\tableaurule
  \kern\normallineskip\hbox
    {\kern\normallineskip\vbox
      {\gettableau#1 0 }%
     \kern\normallineskip\kern\tableaurule}%
  \kern\normallineskip\kern\tableaurule}}
\def\gettableau#1 {\ifnum#1=0\let\next=\null\else
  \squares{#1}\let\next=\gettableau\fi\next}

\tableauside=1.0ex
\tableaurule=0.4pt

\newif\iflanl
\openin 1 lanlmac
\ifeof 1 \lanlfalse \else \lanltrue \fi
\closein 1
\iflanl
    \input lanlmac
\else
    \message{[lanlmac not found - use harvmac instead}
    \input harvmac
    \fi
\newif\ifhypertex
\ifx\hyperdef\UnDeFiNeD
    \hypertexfalse
    \message{[HYPERTEX MODE OFF}
    \def\hyperref#1#2#3#4{#4}
    \def\hyperdef#1#2#3#4{#4}
    \def\hypernoname{}
    \def\e@tf@ur#1{}
    \def\eprt#1{{\tt #1}}
    \def\CERN{\centerline{CERN Theory Division}}
    \def\wl{W.\ Lerche}
\else
    \hypertextrue
    \message{[HYPERTEX MODE ON}
\def\eprt#1{{\tt
#1}}
\def\CERN{\centerline{

CERN Theory Division}}
\def\wl{
 W.\ Lerche}
\fi
\newif\ifdraft

\noblackbox
%
\newif\iffigureexists
\newif\ifepsfloaded
\def\epsfcheck{
\ifdraft
\input epsf\epsfloadedtrue
\else
  \openin 1 epsf
  \ifeof 1 \epsfloadedfalse \else \epsfloadedtrue \fi
  \closein 1
  \ifepsfloaded
    \input epsf
  \else
\immediate\write20{NO EPSF FILE --- FIGURES WILL BE IGNORED}
  \fi
\fi
\def\epsfcheck{}}
\def\checkex#1{
\ifdraft
\figureexistsfalse\immediate%
\write20{Draftmode: figure #1 not included}
\figureexiststrue
\else\relax
    \ifepsfloaded \openin 1 #1
        \ifeof 1
           \figureexistsfalse
  \immediate\write20{FIGURE FILE #1 NOT FOUND}
        \else \figureexiststrue
        \fi \closein 1
    \else \figureexistsfalse
    \fi
\fi}
\def\missbox#1#2{$\vcenter{\hrule
\hbox{\vrule height#1\kern1.truein
\raise.5truein\hbox{#2} \kern1.truein \vrule} \hrule}$}
\def\lfig#1{
\let\labelflag=#1%
\def\numb@rone{#1}%
\ifx\labelflag\UnDeFiNeD%
{\xdef#1{\the\figno}%
\writedef{#1\leftbracket{\the\figno}}%
\global\advance\figno by1%
}\fi{\hyperref{}{figure}{{\numb@rone}}{Fig.{\numb@rone}}}}
\def\figinsert#1#2#3#4{
\epsfcheck\checkex{#4}%
\def\figsize{#3}%
\let\flag=#1\ifx\flag\UnDeFiNeD
{\xdef#1{\the\figno}%
\writedef{#1\leftbracket{\the\figno}}%
\global\advance\figno by1%
}\fi
\goodbreak\midinsert%
\iffigureexists
\centerline{\epsfysize\figsize\epsfbox{#4}}%
\else%
\vskip.05truein
  \ifepsfloaded
  \ifdraft
  \centerline{\missbox\figsize{Draftmode: #4 not included}}%
  \else
  \centerline{\missbox\figsize{#4 not found}}
  \fi
  \else
  \centerline{\missbox\figsize{epsf.tex not found}}
  \fi
\vskip.05truein
\fi%
{\smallskip%
\leftskip 4pc \rightskip 4pc%
\noindent\ninepoint\sl \baselineskip=11pt%
{\bf{\hyperdef\hypernoname{figure}{{#1}}{Fig.{#1}}}:~}#2%
\smallskip}\bigskip\endinsert%
}

\def\boxit#1{\vbox{\hrule\hbox{\vrule\kern8pt
\vbox{\hbox{\kern8pt}\hbox{\vbox{#1}}\hbox{\kern8pt}}
\kern8pt\vrule}\hrule}}
\def\mathboxit#1{\vbox{\hrule\hbox{\vrule\kern8pt\vbox{\kern8pt
\hbox{$\displaystyle #1$}\kern8pt}\kern8pt\vrule}\hrule}}
%
%
\def\abstract#1{
\vskip .5in\vfil\centerline
{\bf Abstract}\penalty1000
{{\smallskip\ifx\answ\bigans\leftskip 1pc \rightskip 1pc
\else\leftskip 1pc \rightskip 1pc\fi
\noindent \abstractfont  \baselineskip=12pt
{#1} \smallskip}}
\penalty-1000}
%
\def\hth/#1#2#3#4#5#6#7{{\tt hep-th/#1#2#3#4#5#6#7}}
\def\nup#1({Nucl.\ Phys.\ $\us {B#1}$\ (}
\def\plt#1({Phys.\ Lett.\ $\us  {B#1}$\ (}
\def\cmp#1({Comm.\ Math.\ Phys.\ $\us  {#1}$\ (}
\def\prp#1({Phys.\ Rep.\ $\us  {#1}$\ (}
\def\prl#1({Phys.\ Rev.\ Lett.\ $\us  {#1}$\ (}
\def\prv#1({Phys.\ Rev.\ $\us  {#1}$\ (}
\def\mpl#1({Mod.\ Phys.\ Let.\ $\us  {A#1}$\ (}
\def\atmp#1({Adv.\ Theor.\ Math.\ Phys.\ $\us  {#1}$\ (}
\def\ijmp#1({Int.\ J.\ Mod.\ Phys.\ $\us{A#1}$\ (}
\def\jhep#1({JHEP\ $\us {#1}$\ (}

\def\bb#1{{\bar{#1}}}
\def\bx#1{{\bf #1}}
\def\cx#1{{\cal #1}}
\def\tx#1{{\tilde{#1}}}

\def\rmx#1{{\rm #1}}
\def\us#1{\underline{#1}}
\def\fc#1#2{{#1\over #2}}
\def\frac#1#2{{#1\over #2}}

\def\br{\hfill\break}
\def\ni{\noindent}

\def\al{\alpha}\def\be{\beta}
\def\eps{\epsilon}\def\om{\omega}

\def\p{\partial}

\def\ZZ{{\bx Z}}
\def\la{\langle}\def\ra{\rangle}

\def\IP{{\bx P}}\def\WP{{\bx {WP}}}
\def\ss{\scriptstyle}

\def\rop{{\cx R}_{\triangle}}
\def\hX{\hat{X}}
\def\QK{F}\def\QKd{F^*}\def\SK{E}\def\SKd{E^*}\def\ss{\scriptstyle}
\def\aa#1{\hfil\!\!{\ss #1}\!\!}
\def\yt#1{\tableau{#1}}
\def\pmut{\underline{P}}
\def\lmut{\underline{L}}\def\rmut{\underline{R}}
\def\bSinf{\{S^a_\infty\}}\def\bRinf{\{R_a^\infty\}}
\def\bRc{\{ {\cx R} _a\} }\def\bSc{\{ {\cx S} ^a\} }

\font\fourteenmi=cmmi10 scaled\magstep 1
\def\tilde{\widetilde}
\def\Trbel#1{\mathop{{\rm Tr}}_{#1}}
\def\Chi{{\hbox{$\textfont1=\fourteenmi \chi$}}}
\def\nihil#1{{\sl #1}}
\def\Hom{{\rm Hom}}
\lref\wip{Work in progress.}
\lref\Witcs{
E.~Witten,
{\it Chern-Simons gauge theory as a string theory},
\eprt{hep-th/9207094}.
}
\lref\rkth{
E.~Witten,
{\it D-branes and K-theory},
JHEP {\bf 9812}, 019 (1998),
\eprt{hep-th/9810188}.
}
\lref\rGH{P. Griffiths and J. Harris, {\it Principles of algebraic
geometry},
Wiley New York 1994.}
\lref\reid{M. Reid, {\it McKay correspondence},
\eprt{alg-geom/9702016};
{\it La correspondance de McKay}, \eprt{math.AG/9911165}.}
\lref\rDD{D.\ Diaconescu and M.\ R.\ Douglas,
\nihil{D-branes on stringy Calabi-Yau manifolds,}
\eprt{hep-th/0006224}.
}\lref\rBott{R.~Bott and L.~W.~Tu,
{\it Differential forms in algebraic topology}, Springer New York
1982.}
\lref\Drev{
M.~R.~Douglas,
{\it Topics in D-geometry},
Class.\ Quant.\ Grav.\ {\bf 17}, 1057 (2000),
\eprt{hep-th/9910170}.}
\lref\rHori{
K.~Hori,
{\it Linear models of supersymmetric D-branes},
\eprt{hep-th/0012179}.
}
\lref\rDBs{{\sl ¬http://www.sctf.net}}
\lref\Dq{
I.~Brunner, M.~R.~Douglas, A.~Lawrence and C.~R\"omelsberger,
{\it D-branes on the quintic},
JHEP {\bf 0008}, 015 (2000), \eprt{hep-th/9906200}.
}
\lref\PM{P.\ Mayr,
\nihil{Phases of supersymmetric D-branes on
K\"ahler manifolds and the McKay correspondence},
JHEP {\bf 0101}, 018 (2001)
\eprt{hep-th/0010223}.
}
\lref\Wlsm{E.\ Witten,
\nihil{Phases of N = 2 theories in two dimensions,}
 Nucl.\ Phys.\ {\bf B403} 159 (1993),
\eprt{hep-th/9301042}.
}
\lref\LVW{
W.~Lerche, C.~Vafa and N.~P.~Warner,
\nihil{Chiral Rings In N=2 Superconformal Theories,}
Nucl.\ Phys.\  {\bf B324} (1989) 427.
}
\lref\IN{Y.~Ito and H.~Nakajima,
\nihil{McKay correspondence and Hilbert schemes in dimension three,}
\eprt{math.AG/9803120}.
}
\lref\Zas{E.\ Zaslow,
\nihil{Solitons and helices: The Search for a math physics bridge,}
 Commun.\ Math.\ Phys.\ {\bf 175} 337 (1996),
\eprt{hep-th/9408133}.
}
\lref\rDR{D.\ Diaconescu and C.\ R\"omelsberger,
\nihil{D-branes and bundles on elliptic fibrations,}
 Nucl.\ Phys.\ B{\bf 574} 245 (2000),
\eprt{hep-th/9910172}.
}
\lref\rKap{M.~M. Kapranov,
{\it On the derived category of coherent
sheaves on Grassmann manifolds}, Math. USSR Izvestiya, {\bf 24} (1985)
183;
{\it On the derived categories of coherent
sheaves on some homogeneous spaces,}
Invent. Math. {\bf 92} (1988) 47.}
\lref\rMut{A.~N.~Rudakov (ed.), ``Helices and vector bundles:
Seminaire Rudakov'', London Mathematical Society Lecture Note Series
148,
Cambridge University Press, Cambridge 1990.}
\lref\Wgm{E.\ Witten,
\nihil{The Verlinde algebra and the cohomology of the Grassmannian,}
\eprt{hep-th/9312104}.
}
\lref\rHil{
H.\ Hiller, \nihil{Geometry of Coxeter groups,}
Pitman London 1982.}
\lref\WLJW{W.\ Lerche and J.\ Walcher,
\nihil{Boundary rings and $N=2$ coset models,}
\eprt{hep-th/0011107}.
}
\lref\DF{M.\ R.\ Douglas and B.\ Fiol,
\nihil{D-branes and discrete torsion.\ II,}
\eprt{hep-th/9903031}.
}
\lref\TM{
A.~Tomasiello,
\nihil{D-branes on Calabi-Yau manifolds and helices,}
JHEP {\bf 0102}, 008 (2001),
\eprt{hep-th/0010217}.
}
\lref\GJ{S.\ Govindarajan and T.\ Jayaraman,
\nihil{D-branes, exceptional sheaves and
quivers on Calabi-Yau manifolds: From Mukai to McKay,}
\eprt{hep-th/0010196}.
}
\lref\MDcat{M.\ R.\ Douglas,
\nihil{D-branes, Categories and N=1 Supersymmetry,}
\eprt{hep-th/0011017}.
}
\lref\HIV{
{K.\ Hori, A.\ Iqbal and C.\ Vafa,
\nihil{D-branes and mirror symmetry,}
\eprt{hep-th/0005247}.
}}
\lref\HV{K.\ Hori and C.\ Vafa,
\nihil{Mirror symmetry,}
\eprt{hep-th/0002222}.
}
\lref\baty{V.\ V.\ Batyrev, I.\ Ciocan-Fontanine,
B.\ Kim and D.\ van Straten,
\nihil{Conifold transitions and mirror symmetry for
Calabi-Yau complete intersections in Grassmannians,}
\eprt{alg-geom/9710022}.
 Nucl.\ Phys.\ {\bf B514} 640 (1998).
}
\lref\EHX{T.\ Eguchi, K.\ Hori and C.\ Xiong,
\nihil{Gravitational quantum cohomology,}
 Int.\ J.\ Mod.\ Phys.\ {\bf A12} 1743 (1997),
\eprt{hep-th/9605225}.
}
\lref\Wsolit{
{P.\ Fendley, S.\ D.\ Mathur, C.\ Vafa and N.\ P.\ Warner,
\nihil{Integrable Deformations And
Scattering Matrices For The $N=2$ Supersymmetric Discrete Series,}
 Phys.\ Lett.\ {\bf B243} 257 (1990);
}
\br
{S.\ Cecotti and C.\ Vafa,
\nihil{On classification of N=2 supersymmetric theories,}
 Commun.\ Math.\ Phys.\ {\bf 158} 569 (1993),
\eprt{hep-th/9211097}.
}
}
\lref\MDGM{M.\ R.\ Douglas and G.\ Moore,
\nihil{D-branes, Quivers, and ALE Instantons,}
\eprt{hep-th/9603167}.
}
\lref\cardy{J.\ L.\ Cardy,
\nihil{Boundary Conditions, Fusion Rules And The Verlinde Formula,}
 Nucl.\ Phys.\ B{\bf 324} 581 (1989).
}

\lref\Balgebra{
{G.\ Pradisi, A.\ Sagnotti and Y.\ S.\ Stanev,
\nihil{Completeness Conditions for Boundary Operators
in 2D Conformal Field Theory,}
 Phys.\ Lett.\ B{\bf 381} 97 (1996),
\eprt{hep-th/9603097};
}
\br
{R.\ E.\ Behrend, P.\ A.\ Pearce, V.\ B.\ Petkova and J.\ Zuber,
\nihil{Boundary conditions in rational conformal field theories,}
 Nucl.\ Phys.\ B{\bf 570} 525 (2000),
\eprt{hep-th/9908036}.
}
}

\lref\pasq{V.\ Pasquier,
\nihil{Operator Content Of The Ade Lattice Models,}
 J.\ Phys.\ AA{\bf 20} 5707 (1987).
}



\vskip-2cm
\Title{\vbox{
\rightline{\vbox{\baselineskip12pt\hbox{CERN-TH/2001-075}
                            \hbox{hep-th/0103114}}}}}
{A New Kind of McKay Correspondence }
\vskip -1.2cm
\centerline{\titlefont From Non-Abelian
Gauge Theories}

\abstractfont
\vskip 1.2cm
\centerline{\wl, P. Mayr
and J. Walcher\foot{Also
Institut f\"ur Theoretische Physik, ETH-H\"onggerberg,
CH-8093 Z\"urich, Switzerland}}
\vskip 0.8cm
\centerline{\CERN}
\centerline{CH-1211 Geneva 23}
\centerline{Switzerland}
\vskip 0.3cm
\abstract{%
The boundary chiral ring of a 2d gauged linear sigma model on a
K\"ahler manifold
$X$ classifies the topological D-brane sectors and the massless open
strings between them. While it is  determined at small volume by simple
group theory, its continuation to generic volume provides
highly non-trivial information about the
$D$-branes on $X$, related to the derived category $D^\flat(X)$.
We use this correspondence to elaborate on an extended notion of
McKay correspondence that captures more general than orbifold
singularities.
As an illustration, we work out this new notion of McKay correspondence
for a class of non-compact Calabi-Yau singularities related to
Grassmannians.
}
\Date{\vbox{\hbox{March 2001}
}}
\goodbreak

\parskip=4pt plus 15pt minus 1pt
\baselineskip=14pt
\leftskip=8pt \rightskip=10pt
%


\def\fff{~}
\def\htopop{\cx H^{top}_{op}}\def\htopcl{\cx H^{top}_{cl}}

\def\lra{\, \longrightarrow\, }
\def\pmut{\underline{P}}
\def\lmut{\underline{L}}\def\rmut{\underline{R}}

\def\CC{{\bx C}}\def\CY{Calabi--Yau\ }
\def\ch{{\rm ch}}\def\td{{\rm td}}\def\Ext{{\rm Ext}}

\def\bR{\{R_a\}}\def\bS{\{S^a\}}

\def\Piv{\vec{\Pi}}\def\Qv{\vec{Q}}

\newsec{Introduction}
The understanding of D-brane dynamics on curved spaces with background
fields is an important question. One aspect of great interest is
to connect the different perturbative descriptions of the D-branes
in different regimes of the moduli space\foot{See \Drev\ for an
introduction and \MDcat\ for a more recent list of references.}.
A simple
description of D-branes arises on manifolds $X$ with small curvatures,
where they may be interpreted in terms of equivalence classes of
vector bundles, or more generally, of coherent
sheaves on $X$. In this region of the moduli space
the appropriate description is in terms of the K-theory
group $K(X)$ \rkth, or for a more refined description,
the derived category $\cx D^\flat(X)$ \MDcat. By a
variation of the K\"ahler volumes of $X$ one may then interpolate
to small volume, or large curvatures, where the geometric picture
is corrected by world-sheet quantum effects.
Here the appropriate description
is in terms of the two-dimensional conformal theory on the
world-sheet of the string. Although the corrections may be large,
one nevertheless expects a close correspondence between the D-brane
objects at small and large volumes,
due to the decoupling hypothesis \Dq.\foot{See
also \HIV.}
It states that the holomorphic objects do not depend on the
K\"ahler volume and thus the string corrections enter only at the
level of stability questions of the holomorphic branes.

It was argued in \PM\ that the holomorphic objects may be
classified by the boundary chiral ring $\rop$
in a two-dimensional gauge theory with
vacuum geometry $X$. Indeed the large volume continuation of the
holomorphic small volume objects generated by $\rop$
turn out to have some rather miraculous properties, as will be
reviewed below. For example, there is a canonical identification of
certain truncated modules of $\rop$ with
free generators for $\cx D^\flat(X)$.
A remarkable aspect is that the geometry in question is the
quantum corrected version, which means that this approach allows to
bypass the usual -- quite sophisticated -- methods of
mirror symmetry
with simple algebraic techniques pertaining
to the boundary chiral ring.

The interpolation to large volume provides
an amazingly simple way to extract highly non-trivial geometrical
structures from the simple group theory of the boundary ring.
This leads naturally to a generalized notion of
``McKay correspondence''\foot{We refer to \refs{\reid,\IN} for
an overview and references on the classical McKay correspondence.}
as the isomorphism (predicted by the
phase picture of the 2d gauge theory)
between the representations of the 2d boundary ring
and their large volume sheaf duals.

These ideas apply to any
dimension and to geometries involving
algebraic constraints, and lead to a construction of such a
correspondence
for any, possibly non-unique or only partial, resolution.
In particular the notion of a discrete group, central to the
original McKay correspondence, is replaced by the continuous gauge
group $H$; e.g. the intersections of the compact homology, which
coincide
with the topological open string index, are determined by the structure
constants of the boundary ring, which is isomorphic to
a truncation of the $H$ representation ring \PM. Whether or
not it is possible (and desirable at all) to reduce $H$ to
a discrete group, as for the
classical McKay correspondence, is of little  relevance in this
context.

The purpose
of this note is to substantiate and test these ideas in the more
general case
where the gauge group is non-Abelian, $H=U(k)$, and the exceptional
divisor of the resolution $X\to \hX$ is a Grassmannian $G_{k,n}$.
The study of these geometries for general $k$ is interesting because
their small-volume structure is not an orbifold, in contrast to
what the traditional notion of McKay correspondence is based upon.
Rather, they correspond to more general quotients, for which the
r\^ole of the discrete orbifold group is played by a
certain subgroup $\Gamma'$ of the continuous gauge group $U(k)$.
In particular
the ``tautological sheaves'' are no longer line bundles as in previous
cases.
These geometries thus provide a good testing ground for the advertised
definition of a generalized McKay correspondence in terms of gauged
linear sigma models, while allowing for comparison with independent
results of mathematicians (most notably Kapranov's \rKap\foot{That
one should be able to derive Kapranov's results via a generalized
McKay correspondence has been conjectured by M.\ Reid.}).

The organization of this note is as follows. In sect. 2 we
give a brief summary of the results of \PM, in particular
how the boundary ring of a certain 2d gauge theory
``generates'' the topological sector $\htopop$ of the open string
Hilbert space\foot{It seems quite possible that this correspondence
may be generalized to the non-topological sector as well.}.
In sect. 3 we determine the boundary ring $\rop$ for a special class
of gauge theories with gauge group $U(k)$. The small volume vacuum
geometry
is a (non-compact) Calabi--Yau quotient singularity $\hX$ obtained in
the
zero size limit of a Grassmannian hypersurface $G_{k,n}$,
embedded as an exceptional divisor in the resolution $X\to\hX$.
We construct two bases $\bR$ and $\bS$
for the topological open string Hilbert space $\htopop$
from $\rop$, and determine the index in terms of a truncation of
the representation ring of $H$.
In sect. 4 we interpolate
to large volume (small curvature) and identify the large volume images
$\bRinf$ and $\bSinf$
of the holomorphic small volume bases as collections of exceptional
sheaves
that generate freely the derived category $\cx D^\flat(X)$.
Specifically,
$\bSinf$ represents the homology $H_{*\, c}(X)$
with compact support and the group theoretical topological index
coincides
with the intersection form on the latter.
In sect.~5 we illustrate these ideas in a
case study, for which we work out the fractional brane content,
the quiver diagram and an alternative description in terms
of a local mirror LG model. In sect.~6 we conclude with some comments
on generalizations. In particular we study D-branes on complete
intersection Calabi-Yau 3-folds in Grassmannians.

\newsec{Boundary chiral rings and the derived category $\cx
D^\flat(X)$}

We start with a review of the correspondence between the
boundary chiral ring $\rop$ and the topological sector of the open
string
Hilbert space and its relation to the McKay correspondence \PM.
The general properties of the boundary chiral ring and
its relation to the derived category will be explored in more depth in
\wip;
here we restrict to a formulation and explanation of the proposal and
outline some general arguments. In brief, the proposal says that the
most basic objects of the topological open string sector on a K\"ahler
manifold $X$, namely the allowed boundary conditions and their massless
open string spectra, are classified by the multiplication
ring of chiral fields of a 2d gauge theory at
the boundary\foot{As will be
explained in \wip, this follows from an isomorphism
between the boundary fusion ring
and the path algebra of a quiver construction, and the isomorphism
between the
derived categories of quiver algebras and of coherent sheaves.
This gives an alternative derivation of the fact \MDcat\ that
the category of topological  D-branes is the derived category.}.
As the topological data are closely related to the derived category
$D^\flat(X)$, this provides also the link between group theory and
geometry that leads to a significant
extension of the notion of ``McKay correspondence''.

\subsec{The boundary chiral ring}
As is well-known, the zero mode sector of the closed string
Hilbert space $\htopcl$ on a K\"ahler manifold $X$ has a topological
structure.
It is described by a deformation of the cohomology ring $H^{k,k}(X)$,
called the quantum cohomology ring. The latter is isomorphic to the
quantum chiral ring $\cx R_{cl}$ \LVW, which is the ring of chiral
primary field operators in the $(2,2)$ super-conformal
2d world-sheet theory of the string without boundaries.

The open string sector is described by world-sheets with spatial
boundaries in the 2d CFT. Physics-wise these correspond to
D-branes in the string theory on $X$ and their
world-volume fields support coherent sheaves
on submanifolds of $X$.
The zero mode sector of the Hilbert space $\htopop$ in the boundary
sector
has again a topological structure: it is isomorphic to the
space of sections of $H^{0,k}(X,V)$ \Witcs,
where $V$ is a coherent sheaf of the form $V=E_a^*\otimes E_b$ on $X$,
where
$E_{a,b}$ are the gauge bundles that couple to the two boundaries of the
open
strings, respectively.

Similar as for the closed string sector, the topological sector
$\htopop$ is isomorphic to a ring $\rop$ of 2d chiral field operators
$\Psi$ \PM.
It may be conveniently described in terms of a gauged linear sigma
model
(GLSM), which is a $(2,2)$ supersymmetric 2d gauge theory \Wlsm.
The ring $\rop$ is the multiplication ring of chiral matter fields of
the GLSM
projected to the boundary and it is accordingly referred to as the {\it
boundary
chiral ring}. In the IR, the GLSM model flows to a super-conformal
fixed
point and the ring $\rop$ flows to a boundary analog
of the chiral bulk ring that describes the algebra of topological open
string
vertex operators.
In complete analogy to the bulk chiral ring \LVW,
the mutual OPE of the elements of the boundary ring is regular,
since the dimensions of the fields are protected by their charges \wip.

An important characteristic of the boundary chiral ring are its
structure constants $N$ which appear 
in the operator product:
\eqn\bring{
(\Psi_\mu)_a^{\ b}\,(\Psi_\nu)_b^{\ c}\ \sim\
\sum_\rho {N_{\mu\,\nu}}^\rho(a,b,c)\, (\Psi_\rho)_a^{\ c}\ ,
}
where $(\Psi_\mu)_a^b$ denotes a boundary field in the $(a,b)$ sector
labeled by $\mu$. As far as fusions are concerned,
the fields $(\Psi_\mu)^a_b$ can be replaced
by semi-positive integral matrices $(A_\mu)_a^b$ which
count the open string zero modes mapping between boundary conditions
$a$ and $b$. Then \bring\ turns into a
topological analog of the familiar rational BCFT
relation \Balgebra\ between the annulus coefficients $A$ and the
Verlinde fusion matrices $N$ 
(for this interpretation, we need to sum over $b$ in \bring); however it is more general in that we do not
need to require a rational CFT. An alternative, and quite interesting
interpretation of \bring\ is as a
matrix representation of the path algebra of the associated quiver theory.

Two essential novelties of the boundary chiral ring $\rop$,
as compared to the bulk chiral ring,  are that:
$i)$ the ring elements carry non-trivial representations
of the gauge group~$H$, $ii)$
there is a $\ZZ_2$ gradation corresponding to bosonic and fermionic
chiral super-multiplets.
The first property will be important for generating elements
of $H^{0,p}(X,V)$ with  non-trivial $V$. In fact, by a well-known
property of the IR limit of the GLSM discussed in sect.~4\fff,
the vacuum bundle $E$ for the $H$ gauge fields is non-trivial and
thus the boundary ring elements in non-trivial $H$ representations
become sections of a non-trivial bundle $E$.

The second property is a consequence
of the partial supersymmetry breaking at the boundary; specifically
the super-multiplets of the left-over supersymmetry at the
boundary have bosonic and fermionic statistics in the
directions tangential and normal to the boundary, respectively
\PM\rHori.
This grading into bosonic and fermionic chiral fields at the boundary
defines two sub-rings $\rop^+$ and $\rop^-$ of $\rop$,
generated by the even and odd generators (not elements)
of $\rop$, respectively. This split gives rise to a construction
of two generating bases for $\htopop$ with remarkable properties.
The claim is\foot{There
is a beautiful explanation of this fact that will be fully explored in
\wip.}
that acting with $\rop^\pm$ on a ground state $\cx O_X(m_\pm)$
of $U(1)$ charge $m_\pm$, one obtains naturally two finite bases
$\bR$ and $\bS$ that generate freely the Hilbert space $\htopop$.
As a first check of this claim one may consider the topological open
string
index for $\htopop$
\eqn\index{
\langle E_a, E_b \rangle\ =\  \mathop{{\rm Tr}}_{ab}(-1)^F\ =\ \sum_k\,
(-1)^k\,
\dim\Ext^k(E_a,E_b).}
The r.h.s. is the natural expression for the index in the small volume
phase,
in that it is entirely determined by the group theoretical data \bring\
of the boundary chiral ring.
In fact, modulo extra degeneracy factors from the global symmetry of
the GLSM, the index is given by
an alternating sum over the structure constants\foot{In the 
present context the indices $\mu,\nu..$
and $a,b..$ run over the same set, moreover
the matrices $A$ coincide
with the structure constants~$N$ (that is, they form the
regular representation of the fusion algebra).} $N$ of the fusion 
ring; these turn out to coincide with the structure constants of
(a truncation of) the representation ring of $H$. From this one can
directly
verify that the intersection form on the bases $\bR$  and $\bS$ are
invertible and that they span the lattice of RR charges of $\htopop$.

The language of the boundary chiral ring is particularly appropriate
to describe the topological D-branes in a phase of the 2d gauge theory
corresponding to ``small volume'' of $X$. In fact, for an appropriate
choice of the GLSM, $X$ is the
resolution $X\to \hX$ of a quotient singularity
$\CC^M/\Gamma'$, with $\Gamma'\subset H$ the
(not necessarily discrete) quotient group. A change of parameters in
the 2d gauge theory then interpolates from $\hX$ to the resolution $X$,
by giving finite K\"ahler volume to the exceptional divisor $E\subset
X$ \Wlsm.

Continuing the holomorphic objects in $\bR$ and $\bS$ to generic volume
in this way leads to the following, remarkable link to the derived
category
$\cx D^\flat(X)$.

\subsec{The derived category $\cx D^\flat(X)$ and McKay}
Let $\bSinf$ and $\bRinf$ denote the generic, ``large'' volume
counter parts of the bases $\bR$ and $\bS$ constructed from the
boundary chiral ring at small volume. They are
expected to be collections of coherent sheaves on $X$,
and, by Hirzebruch-Riemann-Roch, the open string index has now a
natural representation in terms of geometric integrals, i.e.:
\eqn\HRR{(*)\ \
\langle E_a, E_b \rangle=
\sum_k\, (-1)^k\,
\dim\Ext^k(E_a,E_b)\matrix{{\ss HRR}\cr=
\cr{\phantom 1}}
\int_X \ch(E_a^*)\, \ch(E_b)\, \td(X).}
The large volume bases $\bSinf$ and $\bRinf$ enjoy some miraculous
properties,
summarized in the following conjecture \PM:
\vskip 0.3cm

\ni
$\underline{\rm Conjecture}$:\
\it Let $\bR$ ($\bS$) denote a basis of $\chi(X)$
elements $\in\htopop$,
obtained by acting with the sub-ring $\rop^+$ of bosonic
generators ($\rop^-$ of fermionic generators) of $\rop$
on a ground state $\cx O_X(m_+)$ ($\cx O_X(m_-)$). Then:
\item{i)} The continuation of the bases $\bRinf$ and $\bSinf$
to large volume provides two bases of free generators for the derived
category of coherent sheaves
$\cx D^\flat(E)$. For an appropriate choice of a pair $(m_+,m_-)$ of
integers,
they are orthogonal with respect to the inner product (*).

\item{ii)} The sheaves $\bRinf$ have a non-trivial extension to the
non-compact space $X$ and span $K(X)$. The sheaves $\bSinf$ have
compact support on the exceptional divisor $E$ and span the K-theory
group
$K_c(X)$ with compact support.
The relation (*) defines a ``McKay correspondence'' between
the group theoretical data of a quotient group $\Gamma'\subset H$
and the intersections on the compact homology $H_{*\, c}(X)$.

\item{iii)} The collections $\bRinf$ and $\bSinf$ are exceptional and
generate helices $\cx H_R$ and $\cx H_S$ on $E$. The
collection $\bSinf$ is a special mutation $\pmut$ of $\bRinf$.
\rm
\vskip 0.2cm

\ni
Note that the split into the K-theory groups with compact and
non-compact
support, which derives directly from the matter spectrum of the GLSM,
is very much as in the formulation of
the McKay correspondence by Ito and Nakajima \IN.
Moreover the analytic
continuation to large volume equates the group theoretical
tensor products $\langle S^a,S^b \rangle =$ l.h.s.\HRR\
of the small volume phase
with the intersections $\langle S_\infty^a,S^b_\infty \rangle=$
r.h.s.\HRR\
of the compact cohomology of the resolution, similarly as in
the original McKay correspondence.

In this way the interpolation between small and large volume phases
of boundary sector in the 2d gauge theory leads to a
direct relation between the group theory data of a quotient
singularity and the intersections of any
(partial) resolution. This is in the spirit of the
original McKay correspondence, and it agrees with the ideas of the
mathematicians when restricted to those gauge theories that have
at least two Higgs phases, namely one that describes an orbifold
singularity $\hX=\CC^n/\Gamma$ and another that describes a complete
crepant resolution $X$ of $\hX$.

In the following we apply and test the above ideas for
certain 2d gauge theories with gauge groups $H=U(k)$.

\newsec{Boundary rings for $\cx O_G(-n)$ at small-volume}
We will study now the GLSM with gauge group $H$ with a vacuum geometry
given by the non-compact \CY\ $X=\cx O_G(-n)$,
where $G=G_{k,n}$ is the Grassmannian
parametrizing $k$-planes $\Lambda_k$ through the origin of $\CC^n$.
As the first Chern class $c_1(X)$ vanishes, the total space $X$ of
the canonical bundle $\cx O_G(-n)$ is a non-compact
Calabi--Yau manifold of dimension $d=k\cdot k' +1$, where $k'=n-k$.
The Grassmannian $G_{k,n}$ is the exceptional divisor of the blow
up of a singularity $\hX$ reached in the limit of vanishing K\"ahler
class. A more detailed description of this geometry
is included in App.~A. For the geometry under consideration,
$\dim(\htopcl)=\sum_k h_c^{k,k}(X)=\chi_c(X)=({n \atop k})\equiv N$,
where the subscript $c$ refers to the cohomology with compact support.

\subsec{The GLSM for the non-compact \CY $X=\cx O_{G}(-n)$}
The relevant GLSM is a $(2,2)$ supersymmetric 2d gauge theory
with gauge group $H=U(k)$,
$n$ chiral matter super fields $X_\al,\, \al=1,\dots,n$,
in the fundamental representation and one extra super field $P$
that transforms as $(\det M)^{-n}$,
where $M\in U(k)$ acts on the fundamental representation \Wlsm.
The lowest components $x^i_\al,\,
i=1,\dots,k$ and $p$ of the matter super-fields parametrize the
vacuum geometry of the 2d gauge theory. The D-term equations
impose the following constraints on a supersymmetric vacuum:
\eqn\dterms{
D^i_j = \sum_\al x^i_\al \bb x_{j\al}
- \delta^i_j\, (n\, |p|^2+ r)=0.}
Here $r$ is the FI parameter for the $U(1)$ part of the gauge group,
which is the imaginary part of the complexified K\"ahler class
$t=\fc{\theta}{2\pi}+i\, r$.

For all non-zero values of the second term, the constraint
\dterms\ imposes that the $x^i$ are $k$ orthogonal vectors
in $\CC^n$ of norm $n|p^2|+r$. For $r>0$ this norm is strictly non-zero
and after dividing by the gauge transformations in $U(k)$,
the gauge invariant information described by the $k$ $n$-vectors $x^i$
is a $k$-plane $\Lambda_k\subset \CC^n$. This is the GLSM representation
of the Grassmannian $G_{k,n}$
as the symplectic quotient $\CC^{kn}//U(k)$ \Wlsm.
Moreover, as the extra field $p$ transforms
as a coordinate on the fiber of the bundle $X=\cx O_G(-n)$,
the total target space of the GLSM is the large volume phase
of the non-compact \CY $X=\cx O_G(-n)$ of dimension $d$.

For $r<0$ the allowed vev's include also
the set $\{x^i_\al=0\,\ \forall i,\al\}$,
while on the other hand $p$ must be non-zero.
Setting $p$ equal to one by a $U(1)\subset U(k)$ transformation, leaves
a remaining gauge invariance $\Gamma'$ generated by the $U(k)$ matrices
$M$
with $\det(M)^n=1$. The group $\Gamma'$ contains a continuous subgroup
$SU(k)\subset \Gamma'$ which can
be divided out by passing to the Pl\"ucker embedding of $G_{k,n}$. This
is described in more detail in App.~A. The resulting space is the
$d$-dimensional non-compact Calabi--Yau $\hat{Y}$
which is the cone over a system of quadrics in $\IP^{N-1}$, divided
by the action of the discrete group $\Gamma:\, y_k\to \om y_k$,
with $\om^n=1$. Here the $y_k$ denote homogeneous
coordinates on $\IP^{N-1}$. In the original coordinates $x^i_\al$,
this discrete group
$\Gamma \simeq \ZZ_n\subset \Gamma'$ may be described by the sequence
$$
\ZZ_{kn} \to \Gamma\to SU(k),
$$
where $\ZZ_{kn}$ is generated by the
matrix $\Omega=\tx\om\cdot \bx 1_{k\times k}$
with $\tx\om^{kn}=1$; it fulfills $\Omega^{n}\in SU(k)$.
The gauge transformation generated by $\Omega\in U(k)$
acting on the matrix $(x^i_\al)$ from the left may be
alternatively represented as a
discrete $SU(n)$ transformation acting by the matrix $\tx\om\cdot\bf1$
from the right.

We proceed with a study of the boundary topological sector
of the above GLSM. In a first step we consider the ring structure
generated by the super-fields $X^i_\al$. The extra field $P$
does not introduce new sectors and it will be easy to implement it
at the end.

\subsec{The basis $\bR$ from $\rop^+$}
The lowest components $x^i_\al$ of the even super fields $(\Psi^+_\nu)$
at the
boundary are the projections of the bosonic components in
$X^i_\al$. They are in the representation
$(\us k,\us n)$ of $U(k)\times U(n)$, where $U(k)$ refers to the gauge
group,
while the $U(n)$ acts as a global symmetry and has a trivial
connection.
Let $\nu$ denote the vector that specifies a $SU(m)$ Young tableau with
rows of length $\nu_i$ and total number of boxes
$|\nu|\equiv\sum_i\nu_i$. To avoid confusion, note that the
labels $\nu$ in \bring\ are completely
{\it general} labels for the boundary fields
which may be identified with Young tableaus only for the
specific topological boundary fields $(\Psi_\nu^+)$.
We drop also the boundary sector indices $(a,b)$ in the
following,
as the relevant sector will be obvious from the context.

Acting with the ring $\rop^+$ generated by the fields $x^i_\al$
on a ground state
$\nu(R_1)=(0,\dots,0)$ with $U(1)\subset U(k)$ charge $m_+$,
generates ground states $\Phi^+_\nu \in \htopop$ which are in $H=U(k)$
representations labeled by the $SU(k)$ Young tableaus\foot{
The $U(1)$ charge of the state $\Phi_\nu$ is
fixed by $\nu$ and $m_+$ and will not be explicitly written.
In fact the charge $m_+$ of the ground state (0,\dots,0) can be freely
chosen at this point and corresponds to a choice for the
closed string background.} $\nu$.
We assert that a finite basis $\bR$ of generators for
$\htopop$ may be chosen as a
sequence of $\chi_c(X)=N=({n\atop k})$ elements labeled by
Young tableaus $\{\nu\}$ with at most $k'=n-k$ columns and
at most $k$ rows,
\eqn\rdef{
\us{\rop^+}\quad \rightarrow\quad
\bR = \{\Phi^+_{\nu}\,: \ \nu_1\leq n-k,
\ \nu_i=0 \ \rmx{for}\ i>k\}.
}
First note that the elements of $\bR$ will generate
the charge lattice of the twisted RR gauge fields
precisely if the ``intersection form''
$\Chi^+_{ab}=\langle
R_a, R_b \rangle$ defined by the inner product \HRR
\eqn\inprod{
\langle A,B \rangle = \sum_k\, (-1)^k\,
\dim\Ext^k(A,B),}
is non-degenerate. In particular we may express a state $V\in\htopop$
as the formal integral linear combination $V=\sum_a\la V,R_a \ra \,
(\chi^{+})^{-1\, ab} R_b$,
which describes the twisted RR charges of $V$.

We will verify the non-degeneracy of the intersection form for
the basis $\bR$ below. The meaning of the particular
choice of $N$ symmetrizations \rdef\ is that the set $\bR$ will be
orthogonal to the elements of the set $\bS$ constructed from the
fermionic generators in $\rop^-$ in the next section. In the
latter case the truncation to a specific list of $N$ symmetrizations
will be
entirely fixed by the fermionic statistics of the generators.

The intersection form $\Chi^+$ is determined by
counting the maps $\Phi_\nu\to \Phi_{\nu'}$ with the degree $k$
identified
with the fermion number of the map \PM. Let us first count the
maps associated to the single generator $x^i_1$ of $\rop^+$.
As $x^i_1$ is bosonic, the composition of
maps of degree $>1$ must be totally symmetric.
Thus the contribution $\tx \Chi^+_{ab}$
of $x^i_1$ to $\Chi^+_{ab}$ is
\eqn\chibi{
\tx\Chi^+_{ab}=(N_{\sigma_{m_{ab}}})_{\nu_a}^{\ \nu_b},}
where $\nu_a$ ($\nu_b$) denotes the Young tableau that labels the
symmetrization  of $R_a$  ($R_b$)
and $m_{ab}=|\nu^{(b)}|-|\nu^{(a)}|$. Moreover $\sigma_m$
denotes the $m$-th totally symmetric product and
the $(N_\mu)$ are the fusion coefficients \bring\ of the
boundary chiral ring determined by the tensor product decomposition
\eqn\chibie{
\mu\otimes \nu  = \sum_\rho (N_{\mu})_{\nu}^{\ \rho}\, \rho
}
in $U(\infty)\supset U(k)$.
To obtain the full matrix $\Chi^+_{ab}$ we notice that the
totality of maps from the $x^i_\al$ is the composition of the
maps \chibie\ from the individual $x^i_\al$, and thus $\Chi^+_{ab}$
is simply the $n$-th power of $\tx \Chi^+_{ab}$:
\eqn\chibii{
\langle R_a,R_b \rangle= \Chi^+_{ab}=\big(\tx \Chi^{+\, n} \big)_{ab}
=\sum_\mu (N_{\mu})_{\nu_a}^{\ \nu_b}\cdot \rmx{dim}_{U(n)}(\mu).}
The second expression follows from an alternative counting
of the maps of all $x^i_\al$ at the same time. Namely in addition to
the totally symmetric maps there are now also maps corresponding
to Young tableaus with $i\leq k$ boxes anti-symmetrized. The bosonic
statistics of the $x^i_\al$ implies that the symmetrization of the
global $U(n)$ index $\al$ coincides with that of the $U(k)$ index $i$,
and thus the multiplicity of a map of a  $U(k)$ symmetrization defined
by the Young tableau $\mu$ is the dimension of the ``same''
representation
$\mu$ in $U(n)$. In fact, $\Chi^+_{ab}=(\tx \Chi^+_{ab})^n$ can be
easily
seen to coincide with Kapranov's result\foot{A more
readable account is given in ref.\ \Zas.} \rKap\ for the
relative Euler number for sheaves
on Grassmannians; the relation will be explained in the next
section.

That the matrix $\Chi^+_{ab}$ is invertible
will be shown below
by constructing its inverse.
Note that if we order $\bR$ with increasing $|\nu^{(a)}|$,
as we do in the following, then $\Chi^+_{ab}$ will be upper
triangular.

\subsec{The basis $\bS$ from $\rop^-$}
The lowest components $\psi^i_\al$ of the odd super-fields
$(\Psi^-_\nu)$ arise from the projections of the fermions in the
super-fields $X^i_\al$. Acting with ring $\rop^-$ generated by the
fields $\psi^i_\al$ on a ``trivial'' ground state with $U(1)$ charge
$m_-$
generates another set of ground states $\Phi^-_\nu\in \htopop$.
The discussion is similar to the
previous case for $\rop^+$ with two major modifications:
$i$) the fermionic statistics leads to a natural truncation to a
finite basis of $N$ elements; $ii)$ the lower index $\al$ is
no longer a global symmetry index but participates in
a gauge transformation.

The fermionic statistics implies that the symmetrization $\nu$
of the index $\al=1,\dots,n$ is combined with a symmetrization $\nu^*$
of the $U(k)$ index $i=1,\dots,k$, where $\nu^*$ denotes the Young
tableau
transpose to $\nu$. We may thus label the symmetrization by the
Young tableau $\nu$ for $\al$ only. To proceed we note that
the GLSM contains mass terms for $k^2$ out of the $k\cdot n$
fermions $\psi^i_\al$. In fact the $k\cdot k'$ massless
fermions $\psi^i_\al$ are described by the last term of the sequence
\eqn\massterms{
0\to \rmx{End}(V) \to \CC^{n}\times V \to W \to 0,
}
where $V$ is the $k$-dimensional vector space on which $U(k)$ acts
linearly.
A convenient local gauge choice is
$\psi^i_\al=0$ for $\al=1,\dots,k$,
which leaves $k\cdot k'$ fermions $\psi^i_\al,\, \al=k+1,\dots,n$
as the local generators for the ring $\rop^-$.

{}From the above it follows that the action of $\rop^-$
on the state $\nu=(0,\dots,0)$ of $U(1)$ charge $m_-$
generates the $N$ ground states $\Phi^-_\nu$
specified by the $U(n)$ representations
\eqn\sdef{
\us{\rop^-:}\quad \rightarrow \quad
\bS = \{\Phi^-_\nu\, : \ \nu_1\leq k,\, \nu_i=0\ \rmx{for}\ i>n-k\}.
}
Note that these states carry in addition the representations
$\nu^*$ w.r.t. to the $H=U(k)$ gauge symmetry.

The evaluation of the intersection form
$\Chi_{-}^{ab}=\langle S^a,S^b \rangle$ is similar as before.
Specifically, the multiplicities for the maps from a single
fermion $\psi^i_{\al=1}$ and for the totality of maps,
weighted by fermion number, respectively,  are
\eqn\chif{\eqalign{
\tx \Chi_{-}^{ab}&=(-)^{m_{ab}} (N_{\eps_{m_{ab}}})_{\nu_a}^{\
\nu_b},\cr
\Chi_{-}^{ab}&=({\tx \Chi_{-}}^n)^{ab}
=\sum_\mu (-1)^{|\mu|}\,
(N_\mu)_{\nu_a}^{\ \nu_{b}}\cdot \rmx{dim}_{U(n)}(\mu).
}}
Here $\eps_m$ denotes the $m$-th totally
anti-symmetric representation and, as before, $m_{ab}=|\nu_a|-|\nu_b|$.

\subsec{Orthogonality and a relation to the bulk chiral ring $\cx
R_{cl}$}

To show that
the intersection forms $\chibii$ and $\chif$ are non-degenerate,
we establish now the relation
\eqn\orth{
\sum_b \tx \Chi_-^{ab}\, \tx \Chi^+_{bc}=\sum_{\nu_b} (-)^{m_{ab}}
(N_{\eps_{m_{ab}}})_{\nu_a}^{\ \nu_b}\,
(N_{\sigma_{m_{bc}}})_{\nu_b}^{\ \nu_c}\, =\delta_{ac}}
It implies that for a judicious choice of the ground states $S^1$
and $R_1$, the twisted RR charges of the elements in $\bR$ and $\bS$
generated by the action of the rings $\rop^\pm$ are
related by the  linear transformation
$S^{a\, *}=\Chi_-^{ab}R_b$. This implies in turn the
orthogonality relation
\eqn\orthii{
\langle S^{a\, *}, R_b \rangle = \delta ^a_b.}
The significance of this relation for
the construction of the fractional branes was pointed out in \rDD.
The choice of base points for which the above relation is true
is $m_-=-n-m_+$, as will be derived in the geometric phase
below.

The proof of \orth\ for $k=1$ and generalizations to weighted
projective spaces has been given in \PM.  For general $k$ the
relation can be understood by first noting that the
$(N_\mu)$ coincide with the structure constants of the (classical)
cohomology ring, $H^*(G_{k,n})$; in other words,
the $(N_\mu)$ form a matrix representation of
$H^*(G_{k,n})$.  This follows from their definition \chibie\ in
terms of $U(k)$ tensor products, in conjunction with the result of
\Wgm\ that equates the $U(k)$ fusion rules with the cup product of
the (quantum\foot{In the present situation we encounter the classical
cohomology ring, which amounts to truncating the $U(k)$ fusion
coefficients to upper triangular matrices.}) cohomology ring of
the Grassmannians. It is known that the cohomology ring
is generated by the Chern classes $c_i$, $i=1,...,k$, and these
are represented by the matrices $(N_{\sigma_i})$ associated
associated with the fully symmetric representations.
Moreover the normal Chern classes, $\bar c_{i'}$, are associated with
the totally anti-symmetric Young tableaus\foot{For details
see e.g. \rHil, chapter III.} and are represented
by $(-)^{i'}\, (N_{\eps_{i'}})$, $i'=1,...,k'$.
The orthogonality relation \orth\ is therefore nothing but:
$$
(\sum_{i'}(-1)^{i'}N_{\eps_{i'}})\cdot (\sum_{i} N_{\sigma_i})
=\Big(1+\bar c_1+...+\bar c_{k'}\Big)\cdot
\Big(1+c_1+ ... + c_k\Big)\ =\ \bx 1_{k\times k},
$$
It is thus simply a matrix representation of the equation that
states the triviality of the bundle $\SK \oplus \QK=\CC^n$.

\subsec{Relation to $N=2$ coset models}

Note that $\tilde\Chi$ has showed up in previous work
\WLJW, in the context of the $N=2$ super-conformal coset models
based on $G_{k,n}$.  In that work the open string index
\DF\ $\Chi^{ab}_{CFT}\equiv \Trbel{a,b}(-1)^F$ of the coset boundary
states was computed and found to be given in terms of $U(k)$ fusion
coefficients. Choosing a minimal, non-extended basis of the boundary
states for which the fusion coefficients are
upper-triangular matrices, the index coincides with
$\tilde\Chi^{ab}_-$ as given in \chif.  This means that the CFT
intersection index for the $N=2$ superconformal coset models has
a very close relationship to the intersection form for sheaves on
$G_{k,n}$, i.e.,
\eqn\powermap{
\Chi_-\ =\ \big(\Chi_{CFT}\big)^n.}
This expresses a structural isomorphism between the $N=2$ sigma
model on $G_{k,n}$ and the $N=2$ coset model based on $G_{k,n}$,
which has been known since a long time as far as topological bulk
physics is concerned \LVW. That is, the (appropriately perturbed)
chiral rings of these models are isomorphic, even though the charges
of the chiral fields are different.
The identity \powermap\ may be viewed as the reflection of this in
the boundary sector of these models.
It says that the boundary rings are isomorphic up to multiplicities.

\subsec{The extra field $P$}
So far we have neglected the generators in $\rop$
associated to the extra field $P$ that adds the non-compact
direction of the fiber of $\cx K=\cx O_G(-n)$. As the Hilbert space
$\htopop$ is related to the compact part of the
non-compact Calabi-Yau $X$, the field
$P$  does not add new ground states. However $P$ generates new maps
and thus changes the intersection forms $\Chi^+_{ab}$ and
$\Chi_-^{ab}$.
As the field $P$ is associated to the canonical bundle,
the additional maps follow most easily from Serre duality
\eqn\serre{
[\, H^k(X,V)\, ]^* \simeq H^{d-k}(X,V^*\otimes \cx K),}
with the result that
\eqn\xis{
\Chi^+(X)=\Chi^++(-)^n\Chi^{+\, T},\qquad
\Chi_-(X)=\Chi_-+(-)^n\Chi_-^{T} .}
In fact these expressions agree with those derived in \PM\
for the intersection form of the restrictions of the same sheaves to a
compact hypersurface $Y$ embedded in the exceptional divisor of $X$.

\newsec{The large volume phase: exceptional sheaves on $\cx O_G(-n)$}

A variation of the FI parameter $r$ in $\dterms$ interpolates between
different phases of the 2d gauge theory, and in particular
connects the geometric
quotient at small volume continuously to the large volume phase
that describes the resolution of it. In the large volume
phase the matter fields $X^i_\al$ parameterize the exceptional
divisor $E=G_{k,n}$ of the resolution and are acted upon by
the full $U(k)$ group. Moreover the natural
interpretation of the topological ground states in the boundary sector
is in terms of K-theory \rkth.
As stated in sect. 2\fff, the relevant objects are the large volume
duals,
$\bRinf$ and $\bSinf$, of the two bases $\bR$ and $\bS$
constructed from the chiral boundary ring at small volume. We will now
verify the properties of $\bRinf$ and $\bSinf$ as predicted
by the conjecture~1.

\subsec{Identification of the dual bases $\bRinf$ and $\bSinf$}
There are two universal bundles over $G_{k,n}$, namely the
universal sub-bundle
\eqn\defs{
\SK=\{(\Lambda_k,z)\in G_{k,n}\times \CC^n:\ z\in \Lambda_k\},}
which has the $k$ plane $\Lambda_k\in G_{k,n}$ as its fiber, and the
universal quotient bundle $\QK$ defined by the exact sequence
\eqn\ges{
0\lra \SK \lra \CC^n \lra \QK \lra 0.}
Very importantly, the gauge bundle $U(k)$ of the GLSM is identified in
the IR
with the dual $\SK^*$ of the universal sub-bundle \Wgm. Therefore
the fields $x^i_\al$ related to $\rop^+$ are sections of $\SK^*$
on the resolution $X$ and the dual basis $\bRinf$ is the
collection of bundles
\eqn\rinf{
\bRinf=\{\Sigma^\nu\, \SK^*\},}
where $\nu$ runs over the $N$ Young tableaus in eq.\rdef, and
$\Sigma^\nu\, V$ denotes the symmetrization of the
product bundle $V^{\otimes |\nu|} $ defined by the
Young tableau~$\nu$.

As for the second collection $\bSinf$, we may replace \massterms\ by
$$
0\lra \SK\otimes\SK^* \lra \CC^n\otimes \SK^* \lra \QK\otimes \SK^*
\lra 0,
$$
where the last term is in fact the tangent bundle on $G_{k,n}$,
$\Omega^*=\QK\otimes \SK^*$. To determine the large volume continuation
of the
ground states obtained from products of the fermions $\psi^i_\al$
in $\rop^-$, we have to take into account also their non-trivial
representation under the gauge group $U(k)$. In total, the states take
values in the space of sections of wedge products of the tangent bundle
twisted by $\SK^*$, $\bSinf\subset
\wedge^i (\SK^*\otimes \Omega^*)=\wedge^i (\SK^*\otimes \SK\otimes
\QK)$.
A simplification occurs as the relevant space is, by construction, the
subset
generated by the group of global, holomorphic sections. However
$h^0(\SK\otimes \SK^*)=1$ with the single global section corresponding
to the singlet in the tensor decomposition of $\us{k}\otimes \us{k}$.
Therefore the space of ground states generated by the global sections
is
\eqn\sinf{
\bSinf=\{\Sigma^\nu\, \QK\},}
where $\nu$ runs over the $N$ Young tableaus in eq.\sdef.

As an independent check of these identifications\foot{
For Grassmannians $G_{k,n}$ and more general flag manifolds,
the group theoretical and geometric descriptions are also
related by the Bott-Borel-Weil theorem.},
one may use the r.h.s. of
\HRR\ to verify that the intersection forms $\Chi^+$ \chibii\ and
$\Chi_-$
\chif, as determined from the structure constants of the
boundary ring $\rop$, satisfy\foot{In general the expressions
\chibii\ and \chif\ as determined from the structure constants of the
boundary ring, agree with the geometric integrals only after
taking into account the
appearance of additional massless fields in the large volume phase
\PM.}:
\eqn\chicks{
\Chi^+_{ab} = \int_X \ch(R^{\infty\, *}_a)\, \ch(R^\infty_b)\,
\td(X),\qquad
\Chi_-^{ab} = \int_X \ch(S_\infty^{a\, *})\, \ch(S_\infty^b)\, \td(X).
}
With eqs.\rinf, \sinf, we have precisely recovered Kapranov's
exceptional collections
of sheaves \rKap\ on $G_{k,n}$, which represent free generators for
the derived category $\cx D^\flat(G)$. In physics terms this
means that any D-brane on $G_{k,n}$ can be written in terms of a
bounded
complex involving only either the sheaves $R^\infty_a$, or the
$S_\infty^a$.
{}From the above one may also easily see that the orthogonality
relation
\orthii\ will hold for $R_N=S^{N\, *}$ which implies the
previously mentioned condition $m_-=-n-m_+$.

\subsec{Helices of exceptional sheaves on $G_{k,n}$}
It has been observed in \refs{\PM,\TM,\GJ}
that the large volume versions $\bRinf$ and $\bSinf$
of the ``McKay bases'' for
weighted projective spaces represent foundations of a helix
structure on the exceptional divisor $E$ of the resolution $X\to\hX$.
We briefly describe now how these results extend to the
present case (which is granted given the work of
\rKap) and outline the property $ii)$ of the conjecture.
For details on the definitions of a helix and references we refer to
\Zas.
In brief, a helix $\cx H$ of period $N$ is an infinite series of
exceptional sheaves such that $N$ consecutive elements
represent an exceptional collection. An exceptional sheaf $E$ is
defined by $\Ext^0(E,E)=\CC,\ \Ext^k(E,E)=0,\,  k>0$ and an
exceptional collection $\cx E$ is an ordered collection of exceptional
sheaves with $\Ext^k(E_a,E_b)=0$ for all $a\neq b$ and $k$,
except possibly for a single value of $k$ if $a<b$.

That the collections $\bRinf$ and $\bSinf$ in \rinf\ and \sinf\
are exceptional on $G_{k,n}$,
serve as a foundation for a helix structure and provide, respectively,
free generators for $\cx D^b(G)$ has been shown in
\rKap\foot{To be precise, our definition of the objects $S^a$ differs
by a factor $(-1)^k$ from that in \rKap. This is related to the fact
that
a brane obtained from an odd number of fermions is interpreted as an
anti-brane, see e.g \MDcat.}.
In particular the exceptionality is reflected in the upper triangular
form of the matrices $\Chi^+$ and $\Chi-$ and their unit diagonal.

Whereas the definition of the bundles $R^\infty_a$ as tensor products
of the vacuum bundle $\SK$ is canonical, the same is not in general
true
for the dual bundles $S^a$. Of course, in the present case,
the bundles $S^a$ have been more directly identified as certain
powers of the quotient bundle $\QK$.
However to construct the pull-backs of the sheaves $R_a$ and
$S^a$ to the non-compact space $X$, it is essential to look
at a canonical relation between them given in terms of a certain
operation on exceptional collections, the so-called mutations.

A mutation is an operation on two neighbors of an exceptional
collection that produces a new exceptional collection. There
are two possible cases acting as $(E_a,E_{a+1})\to (E_{a+1},\rmut E_a)$
and $(E_{a-1},E_{a})\to (\lmut E_{a},E_a)$, called a right and left
mutation, respectively. We refer to \refs{\rMut\HIV}
for the details on the definitions.
The two bases $\bRinf$ and $\bSinf$ are related by the special
series of right mutations $\pmut$ \refs{\PM,\TM,\GJ}:
\eqn\defp{
\pmut:\ \{R^\infty_1,...,R^\infty_N\} \longrightarrow
\{R^\infty_N,\rmut R_{N-1},...,\rmut^{N-1}R^\infty_1\}=
\{S_\infty^{{N+1-a}\, *}\}.}
In fact
the relations $S_\infty^{a\, *}=\rmut^{N-a}\, R^\infty_a$ derive from
the
tautological sequence \ges. For the foundation with $R^\infty_N=\cx O$,
we have $R^\infty_{N-1}=\det(\SKd)^{-1}\otimes (\wedge^{k-1}\SKd)=\SK$
and the right mutation defined as
$$
0\lra R^\infty_{N-1}\lra \rmx{Hom}(R^\infty_{N-1},R^\infty_N)^*\otimes
R^\infty_N=\CC^n\lra \rmut R^\infty_{N-1}
\lra 0
$$
coincides with \ges. This recovers our previous result
$S_\infty^{N-1\, *}=\det(\QK)^{-1}\otimes (\wedge^{n-k-1}\QKd)=\QK$.

\subsec{McKay bases on the non-compact space $X$ from the helix on $E$}
We complete now the construction of a McKay correspondence by
extending the definition of the sheaves $R_a$ and $S^a$ to the total
space $X$. As we will only use general properties of the helix on $E$,
the results of this section are more general and apply to
any complete resolution $E\to X \to \hX$.
The corresponding sheaf collections on $X$ will be denoted by
$\bRc$ and $\bSc$ and provide, similarly as in \IN, generators for
the K-theory groups $K(X)$ and $K_c(X)$, respectively.

The sheaves $\cx R_a$ will be simply defined as the
pull-backs of $R_a$ by the restriction map $\pi:\, X\to E$,
where e.g. $E=G_{k,n}$ as in the previous sections. Inspired by an
observation of Tomasiello for $E=\IP^n$ \TM,
we will define the $\cx S^a$ through a certain complex on $X$ that
reduces to the mutation $\pmut$ on the compact exceptional divisor $E$.
In fact the main difference between the collections on $E$ and on $X$
is that
there are no exceptional collections on $X$ because of $c_1(X)=0$. 
Indeed a defining
property of a helix of coherent sheaves is $r^{N-1}E_k=E_{k+N}=E_k(\cx
K^*)$,
where $\cx K^*$ is the anti-canonical bundle. Thus the
definition
of the helix collapses for $\cx K^*=\cx O$ and
one expects instead some kind of
{\it cyclic} structure with period $N$.
Indeed Serre duality implies that the sheaves
$\cx R_a$ and $\cx S^a$ on $X$ have the non-zero extension groups:
\eqn\cyexts{
\Ext^k(\pi^*E_a,\pi^*E_b)=\cases{
\CC,&$a=b,\, k=0,d,$\cr
\Ext^0(E_a,E_b),&$a<b,\, k=0,$\cr
\Ext^d(E_a,E_b),&$a>b,\, k=d,$\cr
0,&else.}
}
Starting from $\cx R_a=\pi^* R_a$, and with $\cx R_a(K^*)=\cx
R_a\otimes \cx K^*(E)$, we define the sheaves $\cx S^{a\, *}$ by
the first of the following two, closely related sequences:
\eqn\sex{
\vbox{\offinterlineskip\tabskip=0pt\halign{\strut
$#$~\hfil&$#$~\hfil&
$#$~\hfil&$#$~\hfil&$#$~\hfil&
$#$~\hfil&$#$~\hfil&\hfil~$#$~\hfil\cr
\cx S^{a\, *}:&\cx R_a(K^*)&\to a^{1}_{ba}\cx R_b(K^*)&
\to a^{2}_{ba}\cx R_b(K^*)&\to...&\to a^{N-1}_{ba}\cx R_b(K^*)&
\to \cx R_a(K^*)\cr\cr
0\to&\cx R_a&\to \tx a^{1}_{ba}\cx R_b&
\to \tx a^{2}_{ba}\cx R_b&\to...&\to \tx a^{N-1}_{ba}\cx R_b&
\to \cx R_a(K^*)\to 0.\cr
}}
}
A crucial point is that in the first sequence, the
sum is over the finite set $b=1,...,N$, reflecting the cyclic structure of $X$,
while in the second sequence the sum is over all $b\in \ZZ$,
reflecting the infinite helix structure on $E$. Accordingly, the 
coefficients $a^{m}_{ba}$ and $\tx a^m_{ba}$ are defined as
\eqn\adef{\eqalign{
\tx a^{m}_{ba}&=\cases{\rmx{dim}\, \Hom(\rmut^{m-1}R_a,R_b)&
$b=m+a+1$\cr 0&else},\cr
a^{m}_{ba}&\equiv\cases{\tx a^{m}_{ba} &$\hskip 2.6cm
b>a$\cr\tx a^{m}_{b+N\, a} &\hskip 2.6cm $b<a$}.}}
In particular the $\tx a^m_{ba}$ describe the morphisms on $E$
and with this definition, the second sequence is a pull-back 
to $X$ of the exact sequence for the identity 
$r^{N-1}R_a=R_a\otimes \cx K^*(E)$. On the other hand, the $a^m_{ba}$ 
describe the cyclic structure on the total 
space $X$ induced by the extra Ext's in eq.\cyexts. 

Reducing to $K$-theory classes, one may then use the second sequence
in \sex\ to rewrite
\eqn\sprop{\eqalign{
\cx S^{a\, *}
=\ &\sum_{m=0}^{N}\sum_{b=1}^{N} (-)^{m-1} a^{m}_{ba}\cx R_b(\cx
K^*)\cr
=\ &\sum_{m=0}^{N}\sum_{b=1}^{N} (-)^{m}\, \tx a^{m}_{ba}
(\cx R_b-\cx R_b(\cx K^*))\cr
=\ &\sum_{m=0}^{N}\sum_{b=1}^{N} (-)^{m}\, \tx a^{m}_{ba}R_b
=\rmut^{N-a}R_a\cr
=\ &S^{a\, *}\hskip 2cm,}
}
where we have defined $\tx a^{m}_{ba}=\delta_{ba}$ for $m=0,N$ and
we have used $\cx R_b-\cx R_b(\cx K^*)=\cx R_b|_E=R_b$.
The above relation $\cx S^a=\cx S^a|_E=S^a$ shows that the K-theory 
classes generated by the collection $\{\cx S^a\}$ are in the compact 
K-theory group $K_c(X)$.

Eqs.\sex\ and \adef\ provide a general definition of the collections
$\bRc$ and $\bSc$ on the non-compact space $X$ which is
the generalization of the ``McKay bases'' of Ito and Nakajima \IN.
It is based solely on the helix structure on the exceptional
divisor $E$ and the above argument implies that the collections
$\bRc$ and $\bSc$ generate the K-theory groups $K(X)$ and $K_c(X)$, 
respectively.

\subsec{Monodromies and the D0-brane}
We finish the section with two further comments. The first concerns
the monodromy group $\cx G$ of the naive, complexified K\"ahler moduli
space.
It is a subgroup of the invariance
group of the intersection form $\Chi_-(X)$. E.g. for the
ADE quotient singularities $\CC^2/\Gamma$, the invariance groups
$\cx G$ are the corresponding Weyl groups.
For the quotients $O_{G_{1,n}}(-n)$
the local monodromy group in fact coincides with $\Gamma$,
namely $\cx G=\ZZ_n$;
a similar statement holds for the generalization to weighted
projective spaces. As a consequence, the foundations $\bRinf$ and
$\bSinf$
for weighted projective spaces $\WP^n$ may be generated from a single
monodromy $T_\infty=AT$ on a hypersurface $Y$ embedded in $\WP^n$
\refs{\PM,\GJ},
where $A$ is the monodromy around the
Gepner point and $T$ the conifold monodromy; moreover a certain
power $T_\infty$ generates the shift of the
complexified K\"ahler class by one. The same is not true
for the singularities $\cx O_{G_{k,n}}(-n)$ for $k>1$,
as a consequence of the fact that the monodromy
at small volume is not Abelian. In fact the invariance group of the
intersection form $\Chi_-(X)$ is a subgroup of the Weyl group of
$U(n)$. It would be interesting to obtain a collection of
generating mutations also in this case.

Secondly, let us identify
the most fundamental bound-state of the fractional D-branes $S^a$,
namely the D0-brane, or the class of a point on $X$.
Its K-theory class is given by the linear combination\foot{An
alternative,
in some sense minimal representation, is discussed in App.~B.}
\eqn\dzero{
[D0]=\sum_a r_a S^a, \quad r_a=\rmx{rank} (R_a).
}
{}From the quotient construction of gauge theories described in
\MDGM\ we expect therefore
that the D0 brane may be obtained as a one-dimensional branch of
the moduli space of a $\prod_a U(r_a)$  gauge theory,
with bi-fundamentals in the representations
specified by the intersection form $\Chi_-(X)$.

\newsec{A case study: Quiver and fractional $D$-branes for
the canonical bundle $O_{G_{2,5}}(-5)$}

As an illustration of the above concepts and the effectiveness and
simplicity of the approach, we present the determination of the McKay
bases, or the spectrum of the fractional branes, for the non-compact
Calabi--Yau $O_{G_{2,5}}(-5)$.

The bases $\bR$ and $\bS$ obtained from the boundary ring $\rop$
carry the representations specified in \rdef\ and \sdef, respectively.
The structure constants \chibi\ of the boundary ring then immediately
determine the intersection forms $\tx \Chi_-$ and
$\Chi_-=(\tx\Chi_-)^5$:
\def\aa#1{\hfil\!\!{\ss #1}\!\!}
\eqn\ktl{\tableauside 4pt
\vbox{
\offinterlineskip\tabskip=0pt\halign{\strut
$#$~\hfil\vrule&
\hfil~$#$~&\hfil~$#$~&\hfil~$#$~&\hfil~$#$~&
\hfil~$#$~&\hfil~$#$~&\hfil~$#$~&\hfil~$#$~&
\hfil~$#$~&\hfil~$#$~
\cr
\tx\Chi_-&\cdot&\yt{1}&\yt{1 1}&\yt{2}&\yt{1 1 1}&\yt{2 1}&\yt{2 1
1}&\yt{2 2}&
\yt{2 2 1}&\yt{2 2 2}\cr
\noalign{\hrule}
\cdot&\aa{1}&\aa{-1}&\aa{0}&\aa{1}&\aa{0}&
\aa{0}&\aa{0}&\aa{0}&\aa{0}&\aa{0}\cr
\yt{1}&\aa{0}&\aa{1}&\aa{-1}&\aa{-1}&\aa{0}&\aa{1}&\aa{0}&\aa{0}&\aa{0}&\aa{0}\cr
\yt{1 1}&\aa{
0}&\aa{0}&\aa{1}&\aa{0}&\aa{-1}&\aa{-1}&\aa{1}&\aa{0}&\aa{0}&\aa{0}\cr
\yt{2}&\aa{
0}&\aa{0}&\aa{0}&\aa{1}&\aa{0}&\aa{-1}&\aa{0}&\aa{1}&\aa{0}&\aa{0}\cr
\yt{1 1 1}&\aa{
0}&\aa{0}&\aa{0}&\aa{0}&\aa{1}&\aa{0}&\aa{-1}&\aa{0}&\aa{0}&\aa{0}\cr
\yt{2
1}&\aa{0}&\aa{0}&\aa{0}&\aa{0}&\aa{0}&
\aa{1}&\aa{-1}&\aa{-1}&\aa{1}&\aa{0}\cr
\yt{2 1 1}&\aa{
0}&\aa{0}&\aa{0}&\aa{0}&\aa{0}&\aa{0}&\aa{1}&\aa{0}&\aa{-1}&\aa{0}\cr
\yt{2 2}&\aa{
0}&\aa{0}&\aa{0}&\aa{0}&\aa{0}&\aa{0}&\aa{0}&\aa{1}&\aa{-1}&\aa{1}\cr
\yt{2 2 1}&\aa{
0}&\aa{0}&\aa{0}&\aa{0}&\aa{0}&\aa{0}&\aa{0}&\aa{0}&\aa{1}&\aa{-1}\cr
\yt{2 2 2}&\aa{
0}&\aa{0}&\aa{0}&\aa{0}&\aa{0}&\aa{0}&\aa{0}&\aa{0}&\aa{0}&\aa{1}\cr
}}
,\qquad
\vbox{
\offinterlineskip\tabskip=0pt\halign{\strut
$#$~\hfil\vrule&
\hfil~$#$~&\hfil~$#$~&\hfil~$#$~&\hfil~$#$~&
\hfil~$#$~&\hfil~$#$~&\hfil~$#$~&\hfil~$#$~&
\hfil~$#$~&\hfil~$#$~
\cr
\Chi_-&\cdot&\yt{1}&\yt{1 1}&\yt{2}&\yt{1 1 1}&\yt{2 1}&\yt{2 1
1}&\yt{2 2}&
\yt{2 2 1}&\yt{2 2 2}\cr
\noalign{\hrule}
\cdot&\aa{ 1}&\aa{ -5}&\aa{ 10}&\aa{ 15}&\aa{ -10}&\aa{ -40}&\aa{
45}&\aa{ 50}&\aa{ -75}&\aa{ 50}\cr
\yt{1}&\aa{  0}&\aa{ 1}&\aa{ -5}&\aa{ -5}&\aa{ 10}&\aa{ 25}&\aa{
-50}&\aa{ -40}&\aa{ 95}&\aa{ -75}\cr
\yt{1 1}&\aa{  0}&\aa{ 0}&\aa{ 1}&\aa{ 0}&\aa{ -5}&\aa{ -5}&\aa{
25}&\aa{ 10}&\aa{ -50}&\aa{ 45}\cr
\yt{2}&\aa{  0}&\aa{ 0}&\aa{ 0}&\aa{ 1}&\aa{ 0}&\aa{ -5}&\aa{ 10}&\aa{
15}&\aa{ -40}&\aa{ 50}\cr
\yt{1 1 1}&\aa{  0}&\aa{ 0}&\aa{ 0}&\aa{ 0}&\aa{ 1}&\aa{ 0}&\aa{
-5}&\aa{ 0}&\aa{ 10}&\aa{ -10}\cr
\yt{2 1}&\aa{  0}&\aa{ 0}&\aa{ 0}&\aa{ 0}&\aa{ 0}&\aa{ 1}&\aa{ -5}&\aa{
-5}&\aa{ 25}&\aa{ -40}\cr
\yt{2 1 1}&\aa{  0}&\aa{ 0}&\aa{ 0}&\aa{ 0}&\aa{ 0}&\aa{ 0}&\aa{
1}&\aa{ 0}&\aa{ -5}&\aa{ 10}\cr
\yt{2 2}&\aa{  0}&\aa{ 0}&\aa{ 0}&\aa{ 0}&\aa{ 0}&\aa{ 0}&\aa{ 0}&\aa{
1}&\aa{ -5}&\aa{ 15}\cr
\yt{2 2 1}&\aa{  0}&\aa{ 0}&\aa{ 0}&\aa{ 0}&\aa{ 0}&\aa{ 0}&\aa{
0}&\aa{ 0}&\aa{ 1}&\aa{ -5}\cr
\yt{2 2 2}&\aa{  0}&\aa{ 0}&\aa{ 0}&\aa{ 0}&\aa{ 0}&\aa{ 0}&\aa{
0}&\aa{ 0}
&\aa{ 0}&\aa{ 1}\cr}}}
Inclusion of the $P$ field leads to the anti-symmetrized matrices
$\Chi^+(X)$ and $\Chi_-(X)$ in eq. \xis, respectively.

Upon interpolation, these small-volume boundary ring data
map to the large volume data which have a geometrical interpretation.
Specifically,
the elements of the large volume bases $\bRinf$ and $\bSinf$ are
the sheaves $\Sigma^\nu\, \SK^*$ and $\Sigma^\nu\, \QK$, resp.,
where $\nu$ runs over the same Young tableaus as above.
The Chern character of these bundles may be expressed in terms of
$c(\SK)=1-c_1+c_2$ by the standard formulae \rBott. Using the relations
$c_2^2-3 c_2 c_1^2+c_1^4=0$, $-3 c_1 c_2^2+4 c_2 c_1^3-c_1^5=0$,
the Todd class
$$\eqalign{
\rmx{td}(G)=&\  1+\fc{5}{2} c_1 +\fc{1}{12} (36 c_1^2+c_2)+
\fc{5}{24} c_1 (11 c_1^2+c_2)\cr&+\fc{1}{720} (897 c_1^4+179 c_2 c_1^2-
3 c_2^2)+\fc{1}{96} c_1 (49 c_1^4+18 c_2 c_1^2-c_2^2)\cr&+
\fc{1}{60480} (9848 c_1^6+6029 c_2 c_1^4-746 c_2^2 c_1^2-72 c_2^3)}
$$
and $\int_X c_2^3=1$ one may verify that the integrals \chicks\
agree with \ktl.  This confirms the advertised
correspondence between the small volume (group theoretical) and large
volume ($K$ theoretical) data.

{}From the intersection data we may draw the quiver graph\foot{A very
similar diagram
appeared in ref.~\WLJW\ for boundary states of
a Kazama-Suzuki coset model, the difference only being in the
multiplicities of the links. As explained before, this reflects
the structural isomorphism between
the coset model and the sigma model on the Grassmannian.} in
\lfig\quiver\
associated to the
exceptional collection $\bSinf$, with a node for each
ground state $S^a$ and a link between nodes representing a
fermionic zero mode contributing to the index $\Chi_-$.
The links indicate the basic maps generated by
the fundamental anti-symmetric representations contributing to $\tx
\Chi_-$.
Specifically fat links denote the five maps generated by the sections
$\psi_\al^i$, and
thin ones the fifteen maps generated by
$\psi_{(\al}^{[i}\psi_{\be)}^{j]}$.
Composing these basic maps according
to \chif\ leads to further links in
the diagram, like for example the dashed ones.
Taking the $P$ field into account adds the links with
reversed arrows.\foot{Note that
although we have drawn the nodes in a $\ZZ_5$ symmetric manner, the
links are not $\ZZ_5$ symmetric; this is in contrast to the quiver for
$O_{\IP^4}(-5)$. The location of
the nodes gets a meaning in the $W$-plane of
the mirror LG model discussed in App.~B, where we also describe a
modified quiver with $\ZZ_5$ symmetric links.}
\figinsert\quiver{
Quiver graph associated with the sheaves $\Sigma^{\al} F$ on
$G_{2,5}$}{1.6in}
{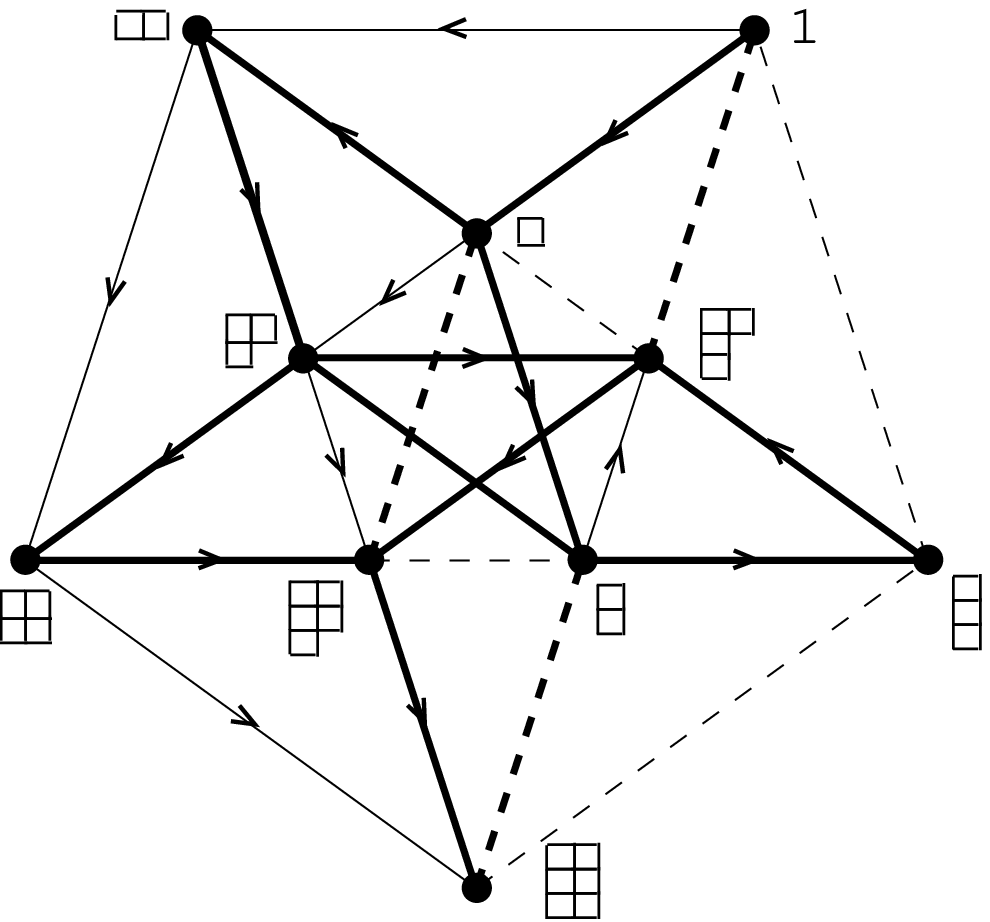}
\vskip -0.7cm
\ni
According to \HIV, intersection diagrams such as the one in
\lfig\quiver\
can also be viewed in terms of solitons of a mirror Landau-Ginzburg
theory. In fact,
a general relation between collections of exceptional sheaves on a Fano
variety and special Lagrangian cycles in a LG theory has been derived
in \HIV\ using local mirror symmetry. This gives another powerful
description of the system of D-branes in terms of the complex
deformations
of a holomorphic superpotential $W$. We include a discussion of these
aspects for the interested reader in App. B.

\newsec{Related quotient singularities and compact Calabi--Yau 3-folds}
The canonical bundle $\cx O_G(-n)$ is just one of a larger
class of non-compact \CY singularities $X$ that share the same
compact homology $H_*(X)=H_*(G_{k,n})$. Specifically the condition
$c_1(X)=0$
is equivalent to the vanishing beta function for the FI parameter
in the GLSM \Wlsm, and thus any
choice of additional matter fields that implies the vanishing of the
beta function leads to a non-compact \CY $X$ with exceptional divisor
$G_{k,n}$\foot{In general,
however, the resolution $X$ will be still singular and may or may not
allow further resolutions to a smooth space.}.

Let us consider only a slight generalization, where the field $P$
is replaced by $m$ fields $P_j$ of $U(1)\subset U(n)$ charges $-q_j$
with
$\sum_j q_j=n$. There is a phase of the gauge theory
where the total space is that of the
bundle $\oplus_j\, \cx O_G(-q_j)$, which is a $k\cdot (n-k)+m$
dimensional
non-compact Calabi--Yau. To describe the quotient singularity,
consider the Pl\"ucker embedding $\IP^N|_{\{Q_i=0\}}$ of $G_{k,n}$,
where
$\{Q_i\}$ is a system of quadrics. The singularity may
be described as follows. Consider space
$\CC^N|_{\{Q_i=0\}}\times_{\CC^*} \CC^m$,
where the $\CC^*$ acts as $(y_i;p_j)\to (\om\, y_i;\om^{-q_j}\, p_j)$
on the coordinates of the two factors. Dividing by $\CC^*$, a
solution of the D-terms for $r<0$ implies that the projection
to the second factor is a $\WP^{m-1}_{\{q_j\}}$. The fiber $F_p$ of
this projection at a point $p\in\WP^{m-1}_{\{q_j\}}$
is $\CC^N/\Gamma$, where $\Gamma \subset U(1)$ is the subgroup
of $U(1)$ that fixes $p$. From the discussion
in sect.\ 3.6, the collections of
ground states $\bR$ and $\bS$ is independent of  a
the choice ``$P$-fields'', however
the intersection forms $\Chi^\pm(X)$ depend on it.

Although the K-theory of these non-compact spaces
might be interesting to study, let us discuss
{\it compact} complete intersection \CY's
$Y\subset G_{k,n}$ defined by the
intersection of the zero locus of the $m$ sections $s_j$ of
$\cx O_G(q_j)$. In the GLSM the constraints are described by the
addition of a
superpotential $\sum_{j=1}^m P_j\, S_j(X^i_\al)$, where
the $S_j$ are the super-fields with $s_j(x^i_\al)$ as lowest
components.
The complex dimension of the CICY $Y$ is then $k\cdot (n-k)-m$.

In particular $m=1,\, q_j=n$ describes the single
hypersurface. Note that in this case
the extension groups in eq.\cyexts\ describe also the
restriction to the hypersurface $Y\subset G \subset X$.
The sheaves $R^\infty_a|_Y$ and $V^a
=S^a_\infty|_Y$
thus give rise to automorphisms, or Fourier-Mukai transforms, on
$K(Y)$:
$$
V \to p_{2*}(p_1^*V\otimes \Delta_{E_a}),\qquad V\in K(Y),\quad
E_a\in\{R^\infty_a|_Y,\, S^a_\infty|_Y,\}
$$
\tableauside7pt
where $\Delta_{E_a}$ is the kernel of the map $E_a
\yt{1}\!\hskip-9pt\times
E_a^* \rightarrow
\cx O_\Delta$ defined on the direct product $Y\times Y$ and
$p_i$ are the projections on its $i$-th factor.

{}From the point of string theory the most interesting case is
dim$_{\CC}(Y)=3$, as $Y$ may then serve as a compactification manifold
for the ten-dimensional string to four dimensions. It turns out
that there are only six choices\foot{We use the notation
$G_{k,n}[q_1,q_2,...,q_m]$ for a complete intersection of $m$
hypersurfaces of degree $q_j$. The mirror maps for the threefolds
in the table have been studied in ref.~\baty.} of integers $(k,n)$
that lead to a ambient space $G_{k,n}$ different from $\IP^{n-1}$:
\eqn\ktl{
\vbox{\offinterlineskip\tabskip=0pt\halign{\strut
$#$~\hfil\vrule&
\hfil~$#$~\hfil&&
\hfil~$#$~\hfil&
\hfil~$#$~\hfil\cr
&\ \chi(Y)\ &&\ \int_Y K^3\ &\ \int_Y c_2\, K\ \cr
\noalign{\hrule}
G_{2,4}[4]      &-176&          &8 &56\cr
G_{2,5}[1,1,3]  &-150&          &15&66\cr
G_{2,5}[1,2,2]  &-120&          &20&68\cr
G_{2,6}[1,1,1,1,2]&-116&        &28&76\cr
G_{2,7}[1,1,1,1,1,1,1]&-98&     &42&84\cr
G_{3,6}[1,1,1,1,1,1]&-96&       &42&84\cr
\noalign{\hrule}}}
}
As in \rDD, D-branes on $Y$ may then be obtained by restriction of the
D-branes on $X$ to $Y$. However the restriction map is {\it not}
good in general as a basis for $\cx D^\flat(X)$ does not necessarily
generate $\cx D^\flat(Y)$. A simple example is the complete
intersection
model $\IP^5[2,4]$ for $Y=G_{2,4}[4]$ obtained from the Pl\"ucker
embedding. It is easy to see that the restriction of the
exceptional collection $\cx O_P(k),\, k=-5,...,0$ for $\IP^5$ to $Y$
generates only a sub-lattice of $H^*(Y)$ by its Chern
classes\foot{A related fact is that
the GLSM with target space $\IP^5[2]$ is not equivalent
to that with target space $G_{2,4}$.}.
Modulo these questions, the calculation of the D-brane spectrum
obtained
by restricting to the complete intersections is straightforward.

As an example\foot{The
results for the other cases are available upon request.}
let us consider the complete intersection $Y=G_{2,5}[1,2,2]$.
The intersection form $\Chi_-(Y)^{ab}=\langle V^a,V^b \rangle_Y$ for
the
restrictions $V^a=S^a|_Y$ is
\def\aa#1{\hfil\!\!{\ss #1}\!\!}
$$
\Chi_-(X)=\pmatrix{\aa{ 0}&\aa{ -5}&\aa{ 10}&\aa{ 16}&\aa{ -9}&\aa{
-40}&\aa{ 40}&\aa{ 52}&\aa{ -65}&\aa{ 38}\cr\aa{  5}&\aa{ 0}&\aa{
-3}&\aa{ -10}&\aa{ 10}&\aa{
16}&\aa{ -49}&\aa{ -30}&\aa{ 92}&\aa{ -65}\cr\aa{  -10}&\aa{ 3}&\aa{
0}&\aa{ 20}&\aa{ -5}&\aa{ -8}&\aa{ 26}&\aa{ 0}&\aa{ -49}&\aa{
40}\cr\aa{  -16}&\aa{ 10}&\aa{ -20
}&\aa{ 0}&\aa{ 2}&\aa{ 80}&\aa{ 0}&\aa{ -104}&\aa{ -30}&\aa{ 52}\cr\aa{
 9}&\aa{ -10}&\aa{ 5}&\aa{ -2}&\aa{ 0}&\aa{ 0}&\aa{ -5}&\aa{ 2}&\aa{
10}&\aa{ -9}\cr\aa{  40}&\aa{ -16}&\aa{
8}&\aa{ -80}&\aa{ 0}&\aa{ 0}&\aa{ -8}&\aa{ 80}&\aa{ 16}&\aa{
-40}\cr\aa{  -40}&\aa{ 49}&\aa{ -26}&\aa{ 0}&\aa{ 5}&\aa{ 8}&\aa{
0}&\aa{ -20}&\aa{ -3}&\aa{ 10}\cr\aa{  -52}&\aa{
30}&\aa{ 0}&\aa{ 104}&\aa{ -2}&\aa{ -80}&\aa{ 20}&\aa{ 0}&\aa{
-10}&\aa{ 16}\cr\aa{  65}&\aa{ -92}&\aa{ 49}&\aa{ 30}&\aa{ -10}&\aa{
-16}&\aa{ 3}&\aa{ 10}&\aa{ 0}&\aa{ -5}\cr
\aa{  -38}&\aa{ 65}&\aa{ -40}&\aa{ -52}&\aa{ 9}&\aa{ 40}&\aa{ -10}&\aa{
-16}&\aa{ 5}&\aa{ 0}}
$$
The rank of this matrix is $\chi(Y)=4$, equal to the number of periods
of $Y$,
and in fact it is easy to verify
that the classes of the $V^a$ generate the K-theory group $K(Y)$ over
the
integers. We may then proceed further and express the fractional branes
$V^a$
in the integral basis of symplectic charges $\Qv$ by a comparison of
the
central charges \Dq\rDR:
\eqn\centralc{
Z(A)=-\int_Y\, e^{-J}\, \ch(A) \, \sqrt{\td(Y)}= \Qv\cdot \Piv,}
where $\Piv=(2\cx F -t\p_tF,\, \p_t \cx F,\, 1,\, t)^T$ is the
period vector of $Y$ with $\cx F$ the prepotential for the special
geometry of the complexified K\"ahler moduli space of $\cx M_Y$.
The polynomial piece of $\cx F$ which fixes the charges $\Qv$ may be
determined from the topological data of the Calabi--Yau $Y$ as in
sect. 9.3. of \PM:
$$
\cx F = -\fc{10}{3} t^3+\fc{17}{6}t +\rmx{const.}\ .
$$
{}From this we obtain the following symplectic charges
$\Qv^a=(Q_6,Q_4,Q_0,Q_2)^T(V^a)$ of the $V^a$:
$$
\Qv^a=
\pmatrix{
\aa{-1}&\aa{ 3}&\aa{ -3}&\aa{ -6}&\aa{ 1}&\aa{ 8}&
\aa{ -3}&\aa{ -6}&\aa{ 3}&\aa{ -1}\cr
\aa{  2}&\aa{ -5}&\aa{ 4}&\aa{ 8}&\aa{ -1}&\aa{ -8}&
\aa{ 2}&\aa{ 4}&\aa{ -1}&\aa{ 0}\cr
\aa{  -38}&\aa{ 65}&\aa{ -40}&\aa{ -52}&\aa{ 9}&
\aa{ 40}&\aa{ -10}&\aa{ -16}&\aa{ 5}&\aa{ 0}\cr
\aa{  -40}&\aa{ 78}&\aa{ -48}&\aa{ -80}&\aa{ 10}&
\aa{ 48}&\aa{ -8}&\aa{ 0}&\aa{ -2}&\aa{ 0}}
$$

\ni {\bf Acknowledgments}:\
We thank M. Reid for email correspondence.

\appendix{A}{Geometry of the non-compact Calabi--Yau $X=\cx O_{G}(-n)$}
We consider the total space of the canonical bundle
$X=\cx O_G(-n)$, where $G=G_{k,n}$ is the Grassmannian
parametrizing $k$-planes $\Lambda_k$ through the origin of $\CC^n$.
The manifold $X$ represents the blow up $X\to \hX$ of a
$d=k(n-k)+1$-dimensional \CY
singularity $\hX$ reached in the limit of vanishing K\"ahler volume of
$G_{k,n}$. E.g. for $k=1$, $\hX=\CC^n/\ZZ_n$ and $G_{1,n}=\IP^{n-1}$ is
the
exceptional divisor of the blow up of $\hX$.

For $k>1$ the \CY $X$ and its singular limit $\hX$  may
be described as follows. The $k$-plane
$\Lambda_k\in \CC^n$ may be represented by $k$ linearly independent
vectors $x^i\in \CC^n$, $i=1,\dots,k$. The $n\cdot k$
components $x^i_\al$, $\al=1,\dots,
n$ provide homogeneous coordinates on $G_{k,n}$ which define local
coordinates after dividing by the group $GL(k)$ that fixes $\Lambda_k$.
The manifold $G_{k,n}$ may be embedded into $\IP^{N-1}$ with
$N=({n\atop k})$ via the global sections of the ample line bundle
$\cx O_G(1)$. More explicitly, the $N$ homogeneous coordinates
$y_{\{r\}}$
of $\IP^N$ are given by the determinants of the $N$ $k\times k$
minors of the $k\times n$ matrix $x^i_\al$ which represent
$N$ global sections of $O(1)$. The image $\varphi(G)\subset \IP^{N-1}$
under this embedding
is given by $({n\atop k+1})$ quadratic relations $Q_i$ of rank $N-d$.
The
embedding $\varphi$ is well-known as the Pl\"ucker embedding \rGH.

For large positive K\"ahler class $\rmx{Im}\, t\gg 0$
of $G_{k,n}$, the image $Y$ of the
total space $X$ is given by $\cx O_{\IP^{N-1}}(-n)$, restricted to the
zero set of the system of quadrics, $\{Q_i=0\}\ \forall i$. To describe
the
image $\hat{Y}=\varphi(\hX)$ of the singularity at small $t$,
we consider the cone $C$ over the set $\{Q_i=0\}\subset \IP^{N-1}$.
Then $C$ is the universal cover of
$\hat{Y}$, which itself is obtained by dividing $C$ by the $\ZZ_n$
action
$y_{\{r\}}\to \om\, y_{\{r\}}$ with $\om^n=1$.

For example,
for $k=1$, $G_{1,n}=\IP^{n-1}$, $x_\al=x^1_\al$ are the $n$ homogeneous
coordinates and the Pl\"ucker embedding is the identity map
$\varphi:\, x_\al\to x_\al$. The singularity $\hat{Y}$ is the
cone over $\IP^{n-1}$, divided by $x_\al \to \om\, x_\al$ with
$\om^{n}=1$.
This is the same as $\CC^{n}/\ZZ_n$ which may also be described as the
cone
over the $n$-th Veronese embedding of $\IP^{n-1}$. For $k=2$, the
embedding $\varphi$ maps the $x^i_\al$ to the homogeneous
coordinates $y_{\al\be}=\eps_{ij}\, x^i_\al x^j_\be$  of $\IP^{N-1}$
with
$N=n(n-1)/2$.
There is one special case where the rank of the quadric system is
maximal
and $\varphi(G)$ is a complete intersection hypersurface, namely
$G_{2,4}$.
In this case the image of under the embedding $\varphi$
is described by the zero locus of the single quadric
$Q:y_{12}y_{34}+y_{13}y_{42}+y_{14}y_{23}=0$ in $\IP^5$.
The image of the total space $\cx O_G(-4)$ is
the cone over $Q$ divided by the $\ZZ_4$ that acts as $y_{\al\be} \to
i y_{\al\be}$ on six coordinates of $\CC^6$.

\appendix{B}{The mirror model for $ G_{2,5}$}
As observed in \PM, the
McKay bases $\bS$ and $\bR$, the orthogonality relation \orthii\
and the mutation $\pmut:\bRinf\to\bSinf$ have a
very transparent interpretation in the $W$-plane of
the mirror model. We restrict to the discussion of the mirror of the
compact divisor $G_{k,n}$ in the following; the superpotential for
the non-compact space obtained as in \refs{\HV,\HIV}\
has the same critical points,
which is consistent with the fact that the non-compact direction does
affect the morphisms between the ground states, but not the
ground states themselves.

Concretely, the mirror of the sigma model on $G_{2,5}$ is (supposedly
\refs{\baty,\HV,\HIV}) described by the Toda potential~\refs\EHX
\eqn\Todapot{
\eqalign{
W\ =&\
{X_1}+{X_1}^{-1}({X_2}+{X_3})+{X_2}^{-1}{X_6}+
({X_2}^{-1}+{X_3}^{-1}){X_4}
\cr
&+({X_6}^{-1}+{X_4}^{-1}){X_5}+{X_5}^{-1}
\ ,}
}
where $X_i$ are coordinates on $(\CC^*)^6$, i.e. we may write
$X_i=e^{-Y_i}$. The critical points of this potential
in the $W$-plane reproduce precisely the nodes of the quiver diagram
in \lfig\quiver; in particular the distances in the diagram have now
a meaning as the 2d masses of solitons that connect
the critical points \Wsolit. Special
Lagrangian cycles correspond to straight
lines in the $W$-plane \HIV.

We have indicated in \lfig\Dzero\ the D-branes\foot{We use here a
labeling in terms of flipped Young tableaus, which directly give
the ranks of the $U(2)$ bundles.} on those SL cycles that are mirror
to the sheaves $\bRinf$ (solid lines) and $\bSinf$ (dashed lines),
respectively.
The latter may be obtained a monodromy that pulls the solid paths
through the critical points such as to obtain the dashed paths.
This monodromy is the image under the mirror
transformation of the mutation $\pmut$ \defp.
In the small volume limit the critical points move to the origin
and the compact SL cycles corresponding to $\bSinf$ collapse. Moreover
the orthogonality relation \orthii\ reflects the presence
of the massless open strings sitting at the nodes, as indicated in the
figure.
\figinsert\Dzero{
Special Lagrangian cycles in the LG mirror of $G_{2,5}$ in the
$W$-plane of \Todapot. The numbers give the fractional brane
content of the $D0$ brane.
 }{1.7in}{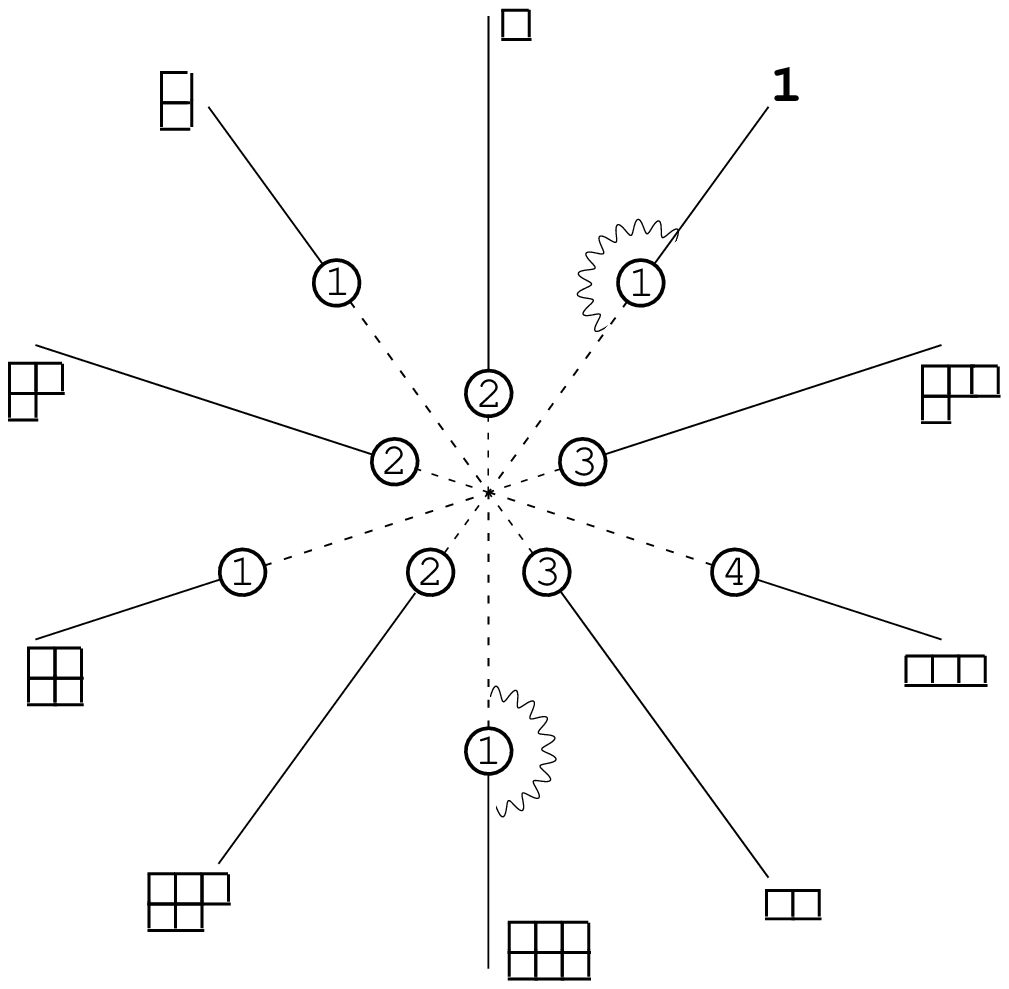}
\def\bRt{\{\tx R_a\}}\def\bSt{\{\tx S^a\}}

\vskip -0.5cm\ni
We have also indicated the fractional brane content of the $D0$ brane
which according to the formula \dzero\ is given by the
ranks of the bundles $R_a$.
The multiplicities of the fractional branes are quite large, and one
may wonder whether there is a more canonical basis of generators $\bRt$
and $\bSt$ for which the multiplicities are smaller. An
exceptional collection that leads to a minimal representation
of the D0 brane may be obtained as follows. We have observed that there
is a mutation of the $\bR$ that leads to a different
exceptional collection $\bRt$ with elements
$$
\tx R_a = \cases{\det(\SK^*)^{\otimes\, (a-1)}&$a$
odd,\cr\det(\SK^*)^{\otimes\, (a-2)}\otimes\SK^*&$a$ even.}
$$
On general grounds \rMut\ the sheaves $\tx R_a$ provide again free
generators
for $\cx D^\flat(X)$.
The canonical ordering induced by the mutation is the series with an
increasing number of boxes. The elements $\tx R_a$ are associated to
the
truncated modules
\tableauside4pt
$(\cdot,\yt{1 1},\yt{2 2},\yt{3 3},\yt{4 4})$ and $(\yt{1},\yt{2
1},\yt{3 2},\yt{4 3},\yt{5 4})$ of the coordinate algebra
$\oplus \Sigma^{(i,i)} \SK^*$; this is similar as in the
algebraic description of $\cx D^\flat(G)$ in \rKap.
Curiously enough, the exchange of Young tableaus generated by this
mutation corresponds precisely to identifications in the
representation ring of $U(2)$ that project to the $U(2)$ fusion ring.
These are of the form  \Wgm\ $V_iU^j=V_{n-2-i}U^{j+n}$, with $U$
the generator of the $U(1)$ and $V_i$ elements of the $SU(2)$
fusion algebra at level $n=5$, associated with the totally symmetric
representations. Specifically, at small radius, the basis $\bRt$
may be obtained by exchanging the ring elements that generate $\bR$
according to the identifications $\yt{2}\to\yt{4 3},\ \yt{3}\to\yt{4
4},\ \yt{3 1} \to\yt{5 4}$ of the $U(2)$ fusion algebra.

In terms of the exceptional collection $\bSt$ defined in K-theory by
$\tx S_\infty^{a\, *}=(\chi(\tx R^\infty_a)^{-1})^{ab} \tx R^\infty_b$
and in the derived category by the sequences \defp, the D0 brane class
takes the following minimal form:
$$
[D0]=1\times \sum_{a=1}^5 \tx S^{2a-1}+2\times \sum_{a=1}^5 \tx S^{2a}\
{}.
$$
This corresponds to a quiver gauge group $U(1)^5\times U(2)^5$; in
the language of \lfig\Dzero, all the outer dots now carry
$\bigcirc\!\hskip-5.3pt 1\,$ while the inner dots carry
$\bigcirc\!\hskip-5.5pt 2$\ .
The change of the total number of component branes due to the mutation
can be understood in terms of brane creation/annihilation
processes induced by the braiding of the critical points of $W$ \HIV.

Another feature of the new basis is that the intersection form on
$X$ becomes manifestly $\ZZ_5$ symmetric.
These considerations may be important for
a construction of the world-volume quiver
theories as a quotient as in \MDGM, which starts from
$r_a^2$ D-branes for each node.

\listrefs
\end